\documentclass[a4paper,11pt]{article}
\pdfoutput=1
\usepackage{jcappub}
\usepackage{aas_macros}
\usepackage{amssymb}
\usepackage{amsmath}
\usepackage{epsfig}
\usepackage[usenames,dvipsnames]{color}
\usepackage{mathrsfs}
\usepackage{graphicx}
\usepackage[export]{adjustbox}
\usepackage{hyperref}

\newcommand{\had}{\textsc{ Had~}}

\newcommand{\onion}  {{\textsc{ Onion}}~}
\newcommand{\onionX}{{\textsc{ Onion}}}

\begin{document}
%%%%%%%%%%%%%%%%%%%%%%%%%%%%%%%

\title{Gamma-radiation sky maps from compact binaries}
\date{\today}

\author[a]{N\'{e}stor Ortiz}
\author[b,c]{Federico Carrasco}
\author[b]{Stephen R. Green}
\author[d]{Luis Lehner}
\author[e]{Steven~L.~Liebling}
\author[f,g]{John Ryan Westernacher-Schneider}

\affiliation[a]{Instituto de Ciencias Nucleares, Universidad Nacional Aut\'onoma de M\'exico,
Circuito Exterior C.U., A.P. 70-543, M\'exico D.F. 04510, M\'exico}
\affiliation[b]{Max Planck Institute for Gravitational Physics (Albert Einstein Institute), 14476 Potsdam, Germany}
\affiliation[c]{ Instituto de F\'i{}sica Enrique Gaviola, CONICET, Ciudad Universitaria, 5000 C\'o{}rdoba, Argentina}
\affiliation[d]{Perimeter Institute for Theoretical Physics, 31 Caroline St., Waterloo, ON, N2L 2Y5, Canada}
\affiliation[e]{Long Island University, Brookville, NY 11548, USA}
\affiliation[f]{Department of Physics \& Astronomy, Clemson University, 118 Kinard Laboratory, Clemson, SC 29634-0978, USA}
\affiliation[g]{Department of Astronomy/Steward Observatory, The University of Arizona, 933 N. Cherry Ave, Tucson, AZ 85721, USA}

\emailAdd{nestor.ortiz@nucleares.unam.mx}
\emailAdd{federico.carrasco@aei.mpg.de}
\emailAdd{stephen.green@aei.mpg.de}
\emailAdd{llehner@perimeterinstitute.ca}
\emailAdd{steve.liebling@liu.edu}
\emailAdd{wester5@clemson.edu}

\abstract{
We study sky maps and light curves of gamma-ray emission from neutron stars in compact binaries, and in isolation. We briefly review some gamma-ray emission models, and reproduce sky maps from a standard isolated pulsar in the Separatrix Layer model. We consider isolated pulsars with several variations of a dipole magnetic field, including superpositions, and predict their gamma-ray emission. Our results provide new heuristics on what can and cannot be inferred about the magnetic field configuration of pulsars from high-energy observations. We find that typical double-peak light curves can be produced by pulsars with significant multipole structure beyond a single dipole.
For binary systems, we also present a simple approximation that is useful for rapid explorations of binary magnetic field structure. Finally, we predict the gamma-ray emission pattern from a compact black hole-neutron star binary moments before merger by applying the Separatrix Layer model to data simulated in full general relativity; we find that face-on observers receive little emission, equatorial observers see one broad peak, and more generic observers typically see two peaks.
}

\maketitle

\tableofcontents

%%%%%%%%%%%%%%%%%%%%%%%%%%%%%%%%%
%%%%%%%%%%%%%%%%%%%%%%%%%%%%%%%%%
\section{Introduction}
%%%%%%%%%%%%%%%%%%%%%%%%%%%%%%%%%
%%%%%%%%%%%%%%%%%%%%%%%%%%%%%%%%%
The binary neutron star (BNS) merger event GW170817 demonstrated the spectacular potential of combined gravitational and electromagnetic observations.
This detection, via gravitational waves (GWs) by the LIGO/Virgo collaboration~\cite{TheLIGOScientific:2017qsa} and electromagnetic~(EM) observations in a multitude of bands 
(e.g. refs.~\cite{Monitor:2017mdv,soares2017electromagnetic,Cowperthwaite:2017dyu,nicholl2017electromagnetic,Chornock:2017sdf,Margutti:2017cjl,alexander2017electromagnetic,Evans:2017mmy,Savchenko:2017ffs}), has provided unprecedented scientific results on the astrophysics of non-vacuum
binary mergers. 
These include new constraints on the equation of state of neutron stars~(NSs)~\cite{Abbott:2018exr,Margalit:2017dij}, improved estimates of the fraction of heavy elements produced in compact binary mergers~\cite{Abbott:2017wuw,Drout:2017ijr,Villar:2017wcc,Nedora:2019jhl}, 
a measurement of the time delay between binary coalescence and the launching of a jet~\cite{geng2019propagation}, a detailed understanding of the off-axis, short-duration gamma-ray burst~(sGRBs)~\cite{Lazzati:2017zsj, Mooley:2017enz, bromberg2018gamma}, and constraints on gravitational theories~\cite{Abbott:2018lct, Baker:2017hug}. 
The scientific impact of this multimessenger detection continues with new observational and theoretical analyses (for recent reviews see refs.~\cite{radice2020dynamics, Dietrich:2020eud, metzger2020kilonovae, Friedman:2020xac}).

Of particular interest is the behavior of binary systems in the late inspiral, merger, and ringdown phases. Studying such phases using GWs alone presents challenges when a NS is involved. These phases occur at high frequencies for BNS and low-mass ratio black hole-neutron star~(BHNS) systems, where current GW detectors lose sensitivity~\cite{read2013matter, maggiore2020science}. Notably, while low-mass ratio BHNS systems can induce the star's disruption, leading to EM
emission due to accretion physics and ejected material (a tantalizing prospect for detection, in light of the recent LIGO/Virgo observations~\cite{Abbott_2021}), high-mass ratio BHNS binaries
would fail to disrupt the star (such as potentially GW190814 that has a mass ratio of roughly 10:1). Since tidal effects are most evident just prior to disruption, high-mass ratio BHNS binaries would be unlikely to leave a tidal imprint on the GW signal that is strong enough to 
distinguish the star from a BH~\cite{Abbott_2020} using current detectors.

To overcome this challenge, EM observations can inform us about the late stages of the merger.
For instance, in the post-merger phase, differences in the characteristics of ejected material,\footnote{Ejection of material (and the accretion disk that can form as a result of the merger of BHNS systems) is more favored when mass ratios are low and/or the BH spin is high} and the corresponding light curves, are used to infer whether a BH promptly forms or whether a massive NS persists for significantly longer than a dynamical time~\cite{Villar:2017wcc, shibata2003merger, shibata2005merger, liu2008general, hotokezaka2011binary, hotokezaka2013mass, dietrich2017gravitational, ruiz2017general,Bernuzzi:2020txg}.
Late pre-merger EM emission can also provide information about
the magnetization of the system~\cite{vietri1996magnetospheric, Hansen:2000am, pshirkov2010radio, mcwilliams2011electromagnetic, lyutikov2011electromagnetic, piro2012magnetic, Lai:2012qe, paschalidis2013general, palenzuela2013electromagnetic, Palenzuela:2013kra, ponce2014interaction, metzger2016pair, lyutikov2019electrodynamics, most2020electromagnetic}.
Such signals can reveal details about the nature of gravity in the relativistic regime close to merger (e.g. refs.~\cite{Hansen:2000am,Lai:2012qe,palenzuela2013electromagnetic,Palenzuela:2013kra,ponce2014interaction}) that would otherwise be inaccessible through GWs alone. The merger phase in particular has the strongest dynamics and curvature effects, which could reveal deviations from  General Relativity~\cite{Damour:1992we,Barausse:2012da,Palenzuela:2013hsa,Sampson:2014qqa}. 

It is thus important to explore the whole gamut of options that can help identify the system's dynamics through any EM band.
On this front, the high-energy EM regime is particularly relevant because the system is expected to be optically thin at such wavelengths~\cite{piran1996gamma}. The prospects for detecting high-energy EM signals would also be enhanced by early warning information from GWs, which could provide timing and localization 
for suitable EM observational facilities to detect them~\cite{cannon2012toward,chan2018binary}.

Two other exciting sets of observations are relevant to our current work.
First, NICER has found an interesting picture of the ``hot spots'' at the NS surface in millisecond pulsars, which would be natural to relate to the stellar magnetic field~\cite{Bogdanov_2019,Bilous_2019}. 
Indeed, these observations suggest a complex magnetic field structure within the NS which could also affect the gamma-ray emission~\cite{Bilous:2019knh,10.1093/mnras/sty2815}. Indeed, on theoretical grounds one might expect an essentially random superposition of dipoles~\cite{thompson1993neutron}. Further, binaries could have an offset effective magnetic dipole, and interactions between binary components might induce global magnetospheric currents associated with magnetic ``twists.'' A twisted magnetic field can then relax as it emits EM radiation, including via dramatic outbursts~\cite{beloborodov2009untwisting, parfrey2013, zrake2016freely, huang2014magnetar, pili2015general, Magnetar}.
We explore both twisted and offset dipole effects in section~\ref{sec:singlestars}.

Second, the astrophysical community has puzzled over observations of fast radio bursts (FRBs) throughout the last decade. While many FRB models have been proposed~\cite{platts2019living, zhang2020physical}, leading models involve a magnetar, with the mechanism often involving a high-energy component of the emission~\cite{popov2007hyperflares, egorov2009possible, popov2010evolution, popov2013millisecond, lyubarsky2014model, Margalit:2018dlt}. In 2020, both ingredients gained strong observational support when an FRB (ST 200428A) was detected in coincidence with high-energy bursts from a known galactic magnetar~\cite{scholz2020bright, bochenek2020independent, 2020NatureFRB1, 2020NatureFRB2}. The high-energy emission pattern from magnetized NSs in general configurations is precisely the subject of the present work.

An important ingredient of NS emission models is the surrounding environment, known as the \emph{magnetosphere}. 
The simplest description for this region is the \textit{vacuum magnetosphere}, in which the EM field is the only source of energy present outside the star. This description however is unable to capture key phenomena, and thus is not considered realistic beyond providing some qualitative guidance for richer models.
A more realistic environment consists of a tenuous plasma, as described in ref.~\cite{Goldreich1969}. The
interaction of the plasma with a rotating star leads to a radiative Poynting flux even in the case where the star's dipole is aligned with its angular momentum~\cite{Spitkovsky:2006np}. A related scenario,
in which a spinning BH interacts with plasma, also leads to collimated Poynting flux, as outlined in ref.~\cite{Blandford1977}. The behavior of the plasma around compact objects can be captured
through the \textit{force-free}~(FF) approximation. Detailed simulations in the FF approximation have enabled the formulation of models that explain observed gamma-ray characteristics, see e.g. refs.~\cite{Bai_2010,contopoulos2010,kalapotharakos2012,kalapotharakos2014,petri2018general}.

In this work, we focus on potential signatures in the gamma-ray band, extrapolating ideas from pulsar theory, for which several emission models have been proposed. These models include the
``Polar Cap''~\cite{sturrock1971model, ruderman1975theory, scharlemann1978potential, Arons:1979bc, harding1978curvature, daugherty1982electromagnetic, sturner1995magnetic}, 
``Outer Gap''~\cite{cheng1986energetic, ho1989spectra, chiang1994outer, 1995ApJ...438..314R, Yadigaroglu-thesis, romani1996gamma, cheng1996general, zhang1997high, hirotani2006particle, hirotani2006high, tang2008revisit, 2000ApJ...537..964C}, 
``Slot Gap''~\cite{1983ApJ...266..215A, Muslimov:2003yz, Muslimov:2004vj, muslimov2004particle}, 
``Two-Pole Caustics''~\cite{Dyks:2003rz, dyks2004two, Dyks:2004ci}, 
``Inner Annular Gap''~\cite{qiao2004inner, qiao2007annular}, 
and
``Separatrix Layer'' models~\cite{Bai_2010}. 
Specific physical effects have been investigated numerically~\cite{contopoulos2010, kalapotharakos2012, kalapotharakos2014} and through first principles approaches~\cite{Cerutti:2015hvk, philippov2018ab, kalapotharakos2018three}. In contrast with gamma-ray or thermal x-ray emission~\cite{Bogdanov_2019,Bilous_2019,lockhart2019x}, our current understanding of pulsar emission in low-energy bands, such as radio, is much less well understood. However, low-energy and high-energy emissions have been associated in theoretical work~\cite{weltevrede2008profile,Cerutti:2015hvk}. 

We thus consider a FF magnetosphere around NSs in several interesting scenarios, and apply the Separatrix Layer~(SL) emission model~\cite{Bai_2010} to predict the angular distribution and relative intensity of their gamma-ray emission. In the SL model, high-energy emission is assumed to occur in a layer just inside the tube of open field lines, producing a caustic effect of photons arriving in phase to the observer. This model generically reproduces the double-peak feature of pulsar gamma-ray light curves~\cite{Bai_2010}. 
As mentioned, the field topology of magnetized systems plays a key role, as does the structure of currents. Such ingredients become increasingly complex as one considers more involved systems, and extending the SL model becomes more difficult. This motivates us to consider isolated neutron star systems with further structure, partly as an intermediate step towards a better understanding of high-energy emission from binaries. This informs our efforts to develop a novel approach to efficiently explore the phenomenology of the magnetosphere in binary systems. This model, the {\it enclosing surface approximation}, consists of a perfectly conducting sphere\footnote{The adoption  of a perfectly conducting sphere to model neutron star magnetospheres became standard practice following the work of ref.~\cite{Spitkovsky:2006np}.} enclosing a hypothetical binary close to merger, and dressed with the magnetic moments of the hypothetical binary. Under the assumptions stated in section~\ref{sec:Enclosing}, this setup is an inexpensive way to develop intuition for the sky maps and light curves of high-energy radiation from BNS magnetospheres.

As a first example, we consider the case of an enclosed offset dipole, which approximates a spin-synchronized binary with a weakly magnetized component. Further examples we consider correspond to (aligned and anti-aligned) enclosed double-dipole superpositions,\footnote{Double-magnetic-dipole superpositions have also been used to model the magnetospheres of single pulsars~\cite{Hamil:2016ylv}.} which approximate binaries of comparably magnetized components.

In section~\ref{sec:BHNS}, we study the emission of a particular BHNS binary using a simulation in full general relativity from ref.~\cite{East:2021spd}.
In contrast with the single pulsar case, the complexity of the binary dynamics requires a more nuanced approach to determine the region of open field lines
from which one expects high-energy emission (see section~\ref{sec:Currents}). We produce and analyze the sky map and light curves from the BHNS system, and
discuss further work that is required to obtain a complete description of the gamma-ray emission.

This article is organized as follows.
In section~\ref{sec:Models}, we briefly review the SL emission model of gamma-radiation from FF pulsar magnetospheres.
We also describe the tools we use to perform numerical simulations of compact objects surrounded by a FF plasma, as well as methods to determine gamma-ray emission zones and directions.
In section~\ref{sec:singlestars}, we present results for single pulsars. We first generate {\it pure dipole} sky maps and light curves, which serve as a fiducial basis for later comparison. We then introduce twists and offsets,
and analyze their deviation from the observable features of the fiducial model as well as robust characteristics largely insensitive to variations considered.
In section~\ref{sec:Binaries}, we generate sky maps of gamma-radiation from the FF magnetosphere of compact binaries, first using the enclosing surface approximation and then in the full generality of a BHNS binary simulation.
We conclude and discuss future directions in section~\ref{sec:Discussion}.

%%%%%%%%%%%%%%%%%%%%%%%%%%%%%%%%%%%%%%%%%%%%%%%%%
%%%%%%%%%%%%%%%%%%%%%%%%%%%%%%%%%%%%%%%%%%%%%%%%%
\section{Models and numerical tools}
\label{sec:Models}
%%%%%%%%%%%%%%%%%%%%%%%%%%%%%%%%%%%%%%%%%%%%%%%%%
%%%%%%%%%%%%%%%%%%%%%%%%%%%%%%%%%%%%%%%%%%%%%%%%%

Predicting the high-energy emission pattern from a magnetized NS involves a number of steps: the numerical simulation of the magnetic field dynamics, the determination of the emission regions, and then the calculation of the emission itself. Here we give a brief overview of the methods and tools used in this work, and provide references for more details. For the application to binary systems, we also discuss shortcomings of the strategy developed for the isolated pulsar scenario. To fix ideas, we find it important to illustrate issues that arise by anticipating results from section~\ref{sec:Binaries}.

%%%%%%%%%%%%%%%%%%%%%%%%
\subsection{Magnetosphere description}
\label{sec:magnetosphere_description}
%%%%%%%%%%%%%%%%%%%%%%%%
We employ two different relativistic magneto-hydrodynamics~(MHD) numerical codes, \had and \onionX, both able to capture the evolution of NSs and their resulting EM fields. Each code has been extensively used in recent
years to explore a number of systems. For the purposes of this work, we provide only a cursory overview of these codes, leaving details to the relevant references below. The reader can skip to section~\ref{sec:emission_model} without loss of continuity. The numerical solutions obtained with these codes were evolved sufficiently long
to remove any transients, and validated via
convergence studies. Each code uses distinct strategies designed for different applications, mainly single vs binary systems. In both cases, given initial data for magnetized stars with a particular field topology, the EM fields evolve under either the FF equations~\cite{FFE} or the resistive MHD (RMHD) equations~\cite{Palenzuela:2012my} until they reach a steady state. Then we extract the EM fields at a particular time and determine the system's high-energy EM radiation under a particular photon emission model.

\had is a distributed numerical infrastructure which uses adaptive mesh refinement~(AMR) on a Cartesian grid to solve a variety of problems. Basic details of \had can be found in the early work of ref.~\cite{lieb}, as well as in more recent studies of BNS mergers (e.g. refs.~\cite{2015PhRvD..92d4045P,Lehner:2016wjg,Sagunski:2017nzb}). To match the ideal MHD environment expected in the dense interior of the NS with a low-density plasma outside, we use RMHD~\cite{Palenzuela:2008sf,Palenzuela:2012my}. This infrastructure has been used to study single stars as well as the mergers of general compact binary systems, for example.
In the current work, we analyze a solution to a BHNS system in full general relativity that was generated with \had and presented previously in ref.~\cite{East:2021spd}.

\onion is a numerical code that solves for the exterior FF (or vacuum) magnetosphere surrounding a NS in a fixed spacetime.
A particular version of the FF equations~\cite{FFE} is evolved using a \textit{multi-block approach}~\cite{Leco_1, Carpenter1994, Carpenter1999, Carpenter2001}, in which the numerical domain is built from several non-overlapping grids where only grid points at their boundaries are shared. In the current work, we make use of flat spacetime in the \onion code to generate single star solutions, as well as to implement the enclosing surface approximation (described in section~\ref{sec:Enclosing}). The code was first described in ref.~\cite{FFE2} for black hole exteriors, and later extended in ref.~\cite{Pulsar} to represent the (assumed) perfectly conducting surface of a NS by means of appropriate boundary conditions at the inner spherical boundary of the computational domain.
As such, no matching is needed between the magnetosphere and the stellar interior, so the code is particularly well suited for scenarios involving a single compact object. In past work, it has been used to generate  accurate solutions in a number
of interesting astrophysical scenarios~\cite{FFE2,Pulsar,Boost,Magnetar,Orbiting, Carrasco:2021jja}.

%%%%%%%%%%%%%%%%%%%%%%%%
\subsection{Emission model}
\label{sec:emission_model}
%%%%%%%%%%%%%%%%%%%%%%%%
Many models of high-energy EM emission from pulsar magnetospheres have been proposed. A key component of such models is the
identification of emission regions. Of particular interest is the separatrix between
closed and open field lines, whose intersection with the star defines the boundaries of the \textit{magnetic polar caps}.  The high-energy emission we model is non-thermal and produced by curvature radiation, synchrotron radiation, 
and inverse Compton scattering from ultra-relativistic particles accelerating in the magnetosphere.
Such particles are assumed to move along open magnetic field lines in a reference frame where the electric field vanishes.

Early emission models (intended to match observed light curves from young pulsars) included the Outer-Gap~(OG)~\cite{cheng1986energetic,chiang1994outer,1995ApJ...438..314R,Yadigaroglu-thesis,2000ApJ...537..964C}, the Slot Gap~(SG)~\cite{1983ApJ...266..215A, Muslimov:2003yz, Muslimov:2004vj, muslimov2004particle} or the Two-Pole Caustics~(TPC) models~\cite{Dyks:2003rz,Dyks:2004ci}. These treated rotating stars with dipolar magnetic fields {\it in vacuum}. In such a ``vacuum magnetosphere'', all field lines are closed. The light cylinder~(LC) is 
aligned with the angular momentum axis and has radius $\rho_\mathrm{LC} = c/\Omega$, where $\Omega$ is the stellar rotational frequency. For modeling emission, field lines which extend beyond the light cylinder are regarded as open, as they are expected to be actually open in more realistic (non-vacuum) scenarios.

Emission models differ primarily in the locations and spatial extents of emission regions.
For instance, the OG model supposes photon emission occurs inside the open field-line tube extending from a null charge surface to the LC, whereas the TPC model assumes emission occurs only along the last open field lines (LOFLs), extending all the way from the NS surface to the LC and even beyond. Despite the success of both the OG and TPC models in reproducing the typical double-peak light curves of young pulsars such as Crab and Vela~\cite{2007ApJ...662.1173H, 2007ApJ...670..677T, 2007ApJ...656.1044T, 10.1111/j.1365-2966.2008.12877.x}, Bai and Spitkovsky revealed large uncertainties in using vacuum fields for predicting gamma-ray light curves~\cite{Bai:2009wf}. The same authors showed that the use of a more realistic, plasma-filled FF magnetosphere leads to robust predictions~\cite{Bai_2010}.\footnote{Although a FF magnetosphere does not formally allow for particle acceleration and photon emission because its electromagnetic field  satisfies ${\bf E} \cdot {\bf B} = 0$ everywhere, it provides a reasonable approximation to the global field structure.} Bai and Spitkovsky introduced the Separatrix Layer (SL) model~\cite{Bai_2010}, in which photon emission occurs within a thin layer sitting on the open side of the surface separating open and closed field lines and extending from the NS surface to a region beyond the LC. In a FF magnetosphere, unlike the vacuum case, field lines are considered open when they actually extend to infinity or close through a \textit{current sheet} (CS). 
For an isolated rotating star with a dipolar magnetic field, such field lines are those extending beyond 
the LC where rotation bends field lines so much that a discontinuity in the magnetic field arises, supported by a CS. For example, when the magnetic dipole is aligned with the rotation axis, the CS lies on the equatorial plane, and has an inner edge at $\rho_{\rm LC}$.
The double peak feature of pulsar light curves is generic in the SL model, where peaks occur due to a ``stagnation" effect, i.e. photon emission from individual field lines arrive to the observer in phase, and is linked to the asymptotic rotating split-monopole structure of the FF field~\cite{Bai_2010}.

Here we implement the SL model under the same assumptions of ref.~\cite{Bai_2010}, namely that radiation is produced by particles accelerated along field lines in the emission zone, with constant emissivity.
Outgoing photons are emitted with a small pitch angle $\theta_\text{p}$ and gyration angle $\varphi$, such that in an orthonormal basis of tangent {\bf t}, normal {\bf n}, and binormal {\bf b} vectors at each point of a magnetic field line in the lab frame, the direction of emission is given by
\begin{equation}
{\bf e} = (A\cos\theta_\text{p}){\bf t} + (A\sin\theta_\text{p}\cos\varphi){\bf n} + (A\sin\theta_\text{p}\sin\varphi){\bf b} + {\bf v}_d,
\end{equation}
where $A$ is determined by the condition that ${\bf e}$ be unitary, and ${\bf v}_d = c\left( {\bf E} \times {\bf B}\right)/B^2$ is the charged particle drift velocity along magnetic field lines in the lab frame. This drifting motion is in addition to their usual gyromotion around magnetic field lines. The local frame moving with the drift velocity ${\bf v}_d$ has vanishing electric field. The observation latitude $\xi_\text{obs}$ and phase of photon emission $\phi$ correspond, respectively, to the polar and (negative) azimuthal angle of ${\bf e}$; see figure~1 in ref.~\cite{Bai:2009wf}.
We take relativistic time delay into account consistently according to ref.~\cite{Bai:2009wf}, so that the phase of emission is shifted by $-{\bf r}\cdot {\bf e}/\rho_\text{LC}$, where ${\bf r}$ is the position vector of light emission. In this work however, we
do not take into account spacetime curvature to compute the photon propagation.
We then produce radiation sky maps, which consist of photon histograms in the $(\phi, \xi_\text{obs})$-space. A light curve then corresponds to a 1-dimensional slice of the radiation sky map at constant observation angle $\xi_\text{obs}$.

%%%%%%%%%%%%%%%%%%%%%%%%%%%%%%%%%%%%%%%%%%%%%%%%
%%%%%%%%%%%%%%%%%%%%%%%%%%%%%%%%%%%%%%%%%%%%%%%
\section{Isolated magnetized neutron stars}
\label{sec:singlestars}
%%%%%%%%%%%%%%%%%%%%%%%%%%%%%%%%%%%%%%%%%%%%%%%
%%%%%%%%%%%%%%%%%%%%%%%%%%%%%%%%%%%%%%%%%%%%%%%
In this section, we study the emission from single stars, deferring binary emission to section~\ref{sec:Binaries}.
To illustrate the validity of our implementations, we first compare our results with those previously
obtained in ref.~\cite{Bai_2010}, which studied the magnetic field of a centered magnetic dipole, possibly misaligned with the rotational axis. 
Then, we introduce a ``twist'' in the magnetosphere and analyze its impact on the resulting high-energy intensity maps. For the configurations studied 
in this work, such a twist does not significantly affect the open field lines, despite constituting a significant perturbation of the star's current. For the purposes of this work, this indicates the degree to which other systems we study would not be affected by twists in closed field lines.
Finally, motivated by observations from the NICER mission as well as non-vacuum
compact binaries, we consider more complex scenarios where the magnetic dipole is offset from the center of the star and possibly misaligned with respect to the rotation axis. 

%%%%%%%%%%%%%%%%%%%%%%%%%%%
\subsection{Pure dipole magnetosphere}
\label{sec:pure}
%%%%%%%%%%%%%%%%%%%%%%%%%%%
We first consider isolated pulsar magnetospheres within the FF approximation.
As typically assumed, we set an initial field configuration given by a centered magnetic dipole in flat spacetime
\begin{eqnarray}\label{eqn:dipole_gral}
\boldsymbol{B}(\boldsymbol{m},\boldsymbol{r}) =  \frac{\mu_0}{4\pi} \left[ 3 \boldsymbol{r} \frac{(\boldsymbol{m}\cdot \boldsymbol{r})}{r^5} - \frac{\boldsymbol{m}}{r^3}\right],
\end{eqnarray}
where $\boldsymbol{m} \equiv m \hat{\boldsymbol{\mu}}$ is the magnetic dipole moment determining the orientation of the magnetic axis. The magnetic dipole vector $\hat{\boldsymbol{\mu}}$ is tilted by an angle $\alpha$ with respect to the rotation axis, which coincides with the $z$-axis of our lab frame.
In all cases, we evolve the system until it relaxes and reaches a stationary state before applying the emission model. Our numerical solutions are found to reach a steady state after
roughly two stellar rotations, and, to be safe, we wait until $3.5$ rotations to extract the FF field
configuration.\footnote{Waiting until $3.5$ rotations contrasts with ref.~\cite{Bai_2010}, where FF fields were extracted at $1.2$ stellar rotations. We note that, for sufficiently long time evolution, polar cap boundaries on misaligned pulsars can form a ``notch'' feature. The formation of a notch could be due to numerical dissipation, or may be a genuine feature of the relaxed magnetosphere. In either case, sky maps and light curves keep their key qualitative features.}

As illustrated in ref.~\cite{Bai_2010}, FF pulsar magnetospheres generically develop CSs that extend beyond the LC. Although CSs ``flap around" in the case of a misaligned dipole, they are otherwise unbroken structures. Strong current layers inside and outside the LC are closely followed by the separatrix layer between open and closed field lines. Therefore, the first step in an empirical method to determine the emission zone for the SL model would be to find the polar cap boundaries 
(the locus of points from which the LOFLs emerge), 
and then ``shrink'' them until most of the emerging lines encounter (or pass close to)
the CS. In order to shrink the cap boundary in a systematic way, it is conventional to parametrize the cap region using open volume coordinates $(\phi_m, r_\text{ov})$, where $\phi_m$ is the magnetic azimuth and $r_\text{ov}$ is the magnetic colatitude~\cite{Yadigaroglu-thesis,Dyks:2004ci}, and then construct rings of constant $r_\text{ov}$. The coordinate $r_\text{ov}$ is chosen such that $r_\text{ov} = 1$ corresponds to the polar cap boundary. For the cases presented here, the separatrix layer typically corresponds to field lines emerging from rings with $ 0.87 \leq r_\text{ov}  \leq 0.92$, depending on the magnetic inclination angle $\alpha$.

Throughout this work, the emission zone consists of a tube centered on field lines emerging from a ring whose $r_\text{ov}$ coordinate we denote by $r_\text{ov}^\text{c}$. Values around $r_\text{ov}^\text{c} \approx 0.9$ were phenomenologically motivated in ref.~\cite{Bai_2010} to generically reproduce the observed double-peaked pulsar light curves using dipole models.
At each side of this central ring, we set up ten rings separated by a constant interval $\Delta r_\text{ov} = 0.005$. Following Dyks {\it et al}~\cite{Dyks:2004ci}, the emissivity decays away from the central ring according to $\exp{\left[ -(r_\text{ov} - r_\text{ov}^\text{c})^2/2\sigma^2 \right]}$ with $\sigma = 0.025$. In order to compare our results to those reported in ref.~\cite{Bai_2010}, we let emission lines extend all the way from the stellar surface up to a cylindrical cutoff radius $\rho_\text{max} = 1.5 \rho_\text{LC}$, we assume a photon pitch angle such that $\sin\theta_\text{p} = 0.1$, and we allow for two gyration angles, $\varphi = 0^\circ, ~180^\circ$, which results in smoother sky maps and more symmetric light curves; both $\theta_\text{p}$ and $\varphi$ remain constant along every emission line.

\begin{figure}[!ht]
\begin{center}
\includegraphics[width=17cm]{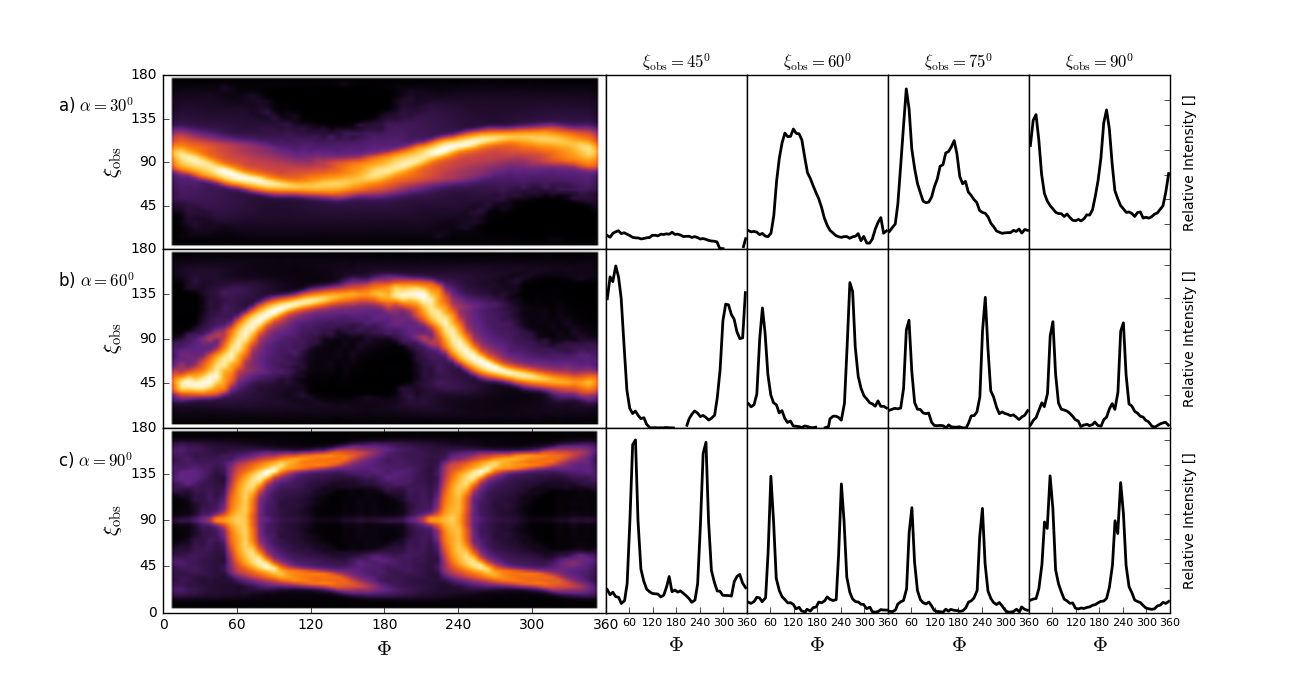}
\end{center}
\caption{\label{fig:Atlas} Sky maps and light curves for a single NS with a FF dipolar magnetosphere (discussed in section~\ref{sec:pure}), all displayed on a linear scale. Three magnetic inclination angles $\alpha$ and four different viewing angles $\xi_\text{obs}$ are displayed. Brighter regions in the sky map correspond to more intense photon emission. For $\alpha = 30^\circ$, the central open volume coordinate is $r^c_\text{ov} = 0.87$; for $\alpha = 60^\circ$, $r^c_\text{ov} = 0.9$; and $r^c_\text{ov} = 0.92$ for $\alpha = 90^\circ$. These results are in excellent agreement with the those reported in ref.~\cite{Bai_2010}.
}
\end{figure}

The atlas in figure~\ref{fig:Atlas} is a collection of sky maps and light curves for representative inclination angles $\alpha \in \{30^\circ, 60^\circ, 90^\circ\}$ and observation latitudes $\xi_\text{obs} \in \{45^\circ, 60^\circ, 75^\circ, 90^\circ\}$. In order to compare with ref.~\cite{Bai_2010}, the rotation rate is chosen such that $R/\rho_{\rm LC} = 0.2$, where $R$ is the stellar radius.
As anticipated in section~\ref{sec:emission_model}, most of the light curves in figure~\ref{fig:Atlas} show two bright and narrow peaks (each magnetic pole contributes one peak), especially as $\alpha$ and $\xi_\text{obs}$ increase.
The presence of these two peaks is a robust, generic feature of the SL model, which is consistent with the typical double-peak feature in light curves from young pulsars. 
These peaks in the light curves are associated with strong caustics in the sky map, due to an effect called ``sky map stagnation" in ref.~\cite{Bai_2010}, which is a generic effect in the sky maps of FF magnetospheres. 
Instead of the accumulation of emission from different field lines, this effect arises from photons emitted from individual field lines arriving in the same region of the sky map. As such, the intensity of the caustics is proportional to the length of particle trajectories in the emission tube, because we assume a constant emissivity.
Therefore, a larger cut-off radius $\rho_\text{max}$ would translate into more intense caustics. Nevertheless, as stated above, in this work we always use $\rho_\text{max} = 1.5 \rho_\text{LC}$.

\begin{figure}[!ht]
\begin{center}
\includegraphics[width=8.0cm]{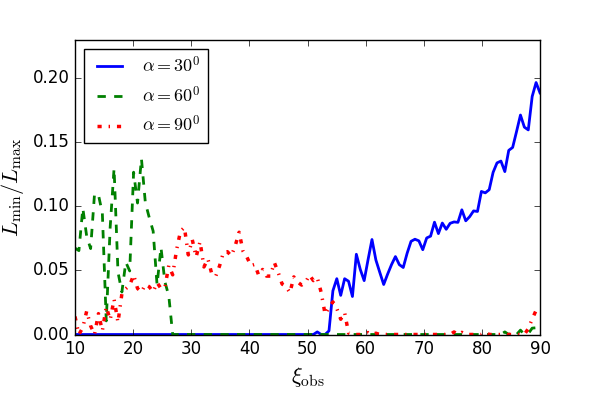}
\end{center}
\caption{\label{fig:pf_pulsars} The ratio $q(\xi_\text{obs})\equiv L_\text{min}/L_\text{max}$ as a function of the viewing angle $\xi_\text{obs}$ corresponding to the pulsars of figure~\ref{fig:Atlas}.
Note the strong dependence of this ratio on the inclination angle $\alpha$ of the dipole.
}
\end{figure}

Given a high-energy EM detection from a pulsar, it is helpful to know the dynamic range of luminosity relative to the maximum. This would indicate
what portion of the emission might lie above a given detector's noise threshold. To that end, we define the dimensionless ratio $q(\xi_\text{obs}) \equiv L_\text{min}/L_\text{max}$, where $L_\text{max(min)}$ is the maximum (minimum) of the luminosity at constant observation
angle $\xi_\text{obs}$.
The ratio $q$ computed for the single dipoles of figure~\ref{fig:Atlas} is plotted as a function of $\xi_\text{obs}$ in figure~\ref{fig:pf_pulsars}, which illustrates a significant minimum contrast ($q \lesssim 0.2$) and dependence on the dipole inclination. Although we do not show plots of $q$ for other cases, we find, in general, significant minimum contrast where $q \lesssim 0.33$ and as low as $q \lesssim 0.04$ in a range of observation angles---with the exception of the orbiting NS described in section~\ref{sec:Enclosing_1dipole}, where $q \lesssim 0.55$.
%

%%%%%%%%%%%%%%%%%%%%%%%%
\subsection{Twisted dipole magnetosphere}
\label{sec:twisted}
%%%%%%%%%%%%%%%%%%%%%%%%
One interesting departure from the pure dipole scenario is to consider a non-potential magnetic field\footnote{That is, a magnetic field which does not derive from a vector potential.} anchored to the NS surface. 
Such configurations, which can sustain global magnetospheric currents emerging and returning to the stellar surface, may arise from different physical processes in the NS interior. 
For instance, it has been suggested that extreme magnetic stresses at the surface of magnetars can lead to crustal deformations or even crustal failures (see e.g. refs.~\cite{Kaspi2017,Thompson_2017}), inducing shear perturbations on the surface magnetic field that propagate into the magnetosphere. A twisted magnetosphere might also be induced in BNS or BHNS systems; since the binary is not tidally locked, significant twisting of field lines can arise, and the resulting built-up tension can then be released through episodic 
reconnection events~\cite{palenzuela2013electromagnetic,ponce2014interaction,Lehner:2011aa}.
 
 A simple model for \textit{twisted magnetospheres} was
used to explain outburst events in magnetars 
that occur when the accumulated twist exceeds a certain critical value. However, below this critical twist, the injected current-carrying bundles are stable within the surrounding FF environment~\cite{parfrey2013, Magnetar}.
We consider this simple model for the stable magnetospheric currents, and analyze potential effects on the resulting gamma-ray sky maps by comparing with the twist-free field configurations in figure~\ref{fig:Atlas}.
 
\begin{figure}[!ht]
\begin{center}
\includegraphics[height=5.5cm,frame]{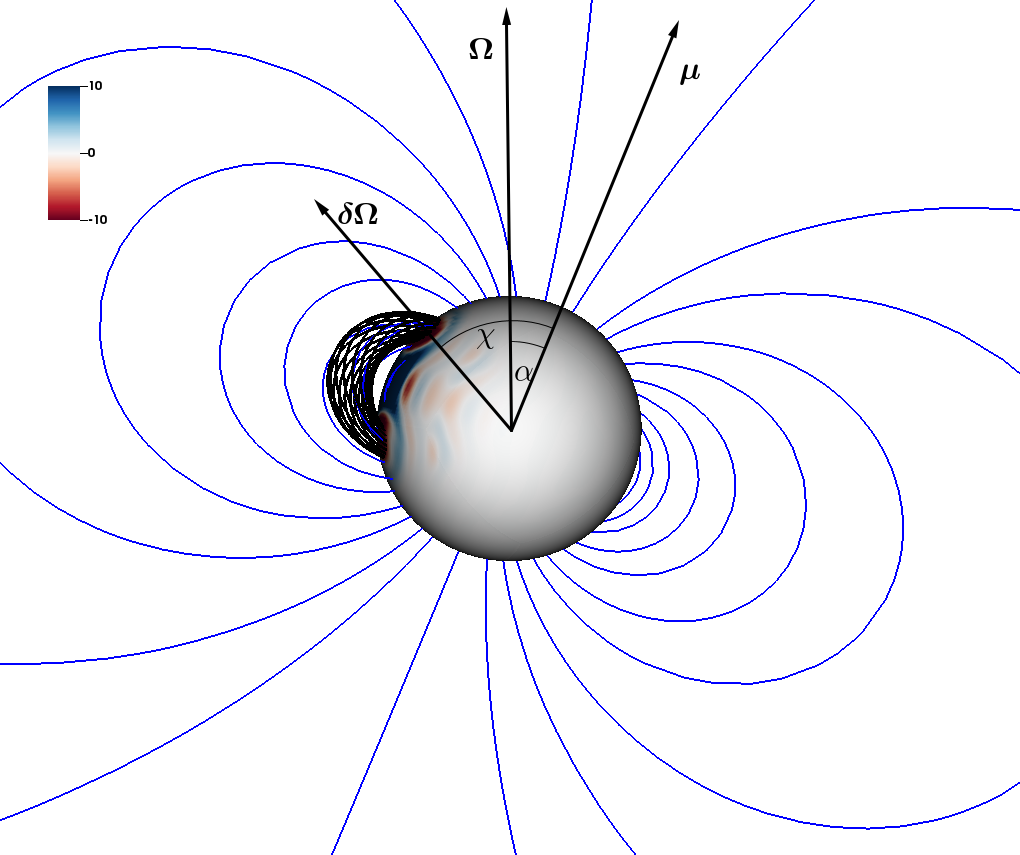}
\end{center}
\caption{Illustration of the dipole magnetosphere with an initialized twist profile described by eq.~\eqref{eq:pc}, before turning on the stellar rotation. The axes of rotation, magnetic dipole, and twist footprint are displayed and labeled with $\Omega$, $\mu$, and $\delta\Omega$, respectively. The red/blue color map on the star represents the electric currents parallel to the magnetic field at the surface (normalized with respect to $\Omega B / 2\pi$), and indicates that the implanted current is about 10 times larger than the current produced by rotation later on.
}
\label{Fig:twist}
\end{figure}

First, we inject magnetic torsion into ``footprints'' on the stellar surface, and then we set the NS (with the stable twisted configuration) into rotation.  
The injected twist is described by a localized rotational perturbation $\delta \Omega$ defined by
 \begin{equation}\label{eq:pc}
 \delta\Omega(\theta^{\prime}, t) = \frac{\omega(t)}{1+\exp \left[\kappa (\theta^{\prime} - \theta^{\prime}_{s}) \right] } 
 \end{equation}
where $\theta^{\prime}_{s}$ is the angular extension of the circular shearing surface from its center at $\theta^{\prime} = 0$, $\omega(t)$ controls the overall twist amplitude over time, and $\kappa$ controls the twist wavelength inside the shearing footprint.
This prescription twists the field lines within the footprints around a chosen axis; the twisted field lines are visible in figure~\ref{Fig:twist}, and the twist axis ($\theta^\prime=0$) is labeled in the figure with $\delta \Omega$. The rotational axis of the star is labeled by $\Omega$ and the magnetic axis by $\mu$. The twist axis forms an angle $\chi$ with respect to the magnetic axis, whereas $\alpha$ again measures the angle between the magnetic and rotational axes. 

The model has a large parameter space, and here we focus on the case with
parameters $\chi/\pi = 0.35 $, $\theta^{\prime}_{s} / \pi = 0.075$ and  $\kappa=30$.
The perturbation amplitude $\omega(t)$ is gradually brought to its maximal value $\omega_o =  10^{-2}$ and then back to zero, leaving an accumulated (maximum) twist of $0.2$ radians. For this configuration, we choose $R/\rho_\mathrm{LC} = 0.1$, so that the twisted bundle lays well-inside the zone of closed field lines (see figure~\ref{Fig:twist}). As anticipated, when the system is set into rotation, the bundle is stable and simply co-rotates with the pulsar. 

\begin{figure}[!ht]
\begin{center}
\includegraphics[width=17cm]{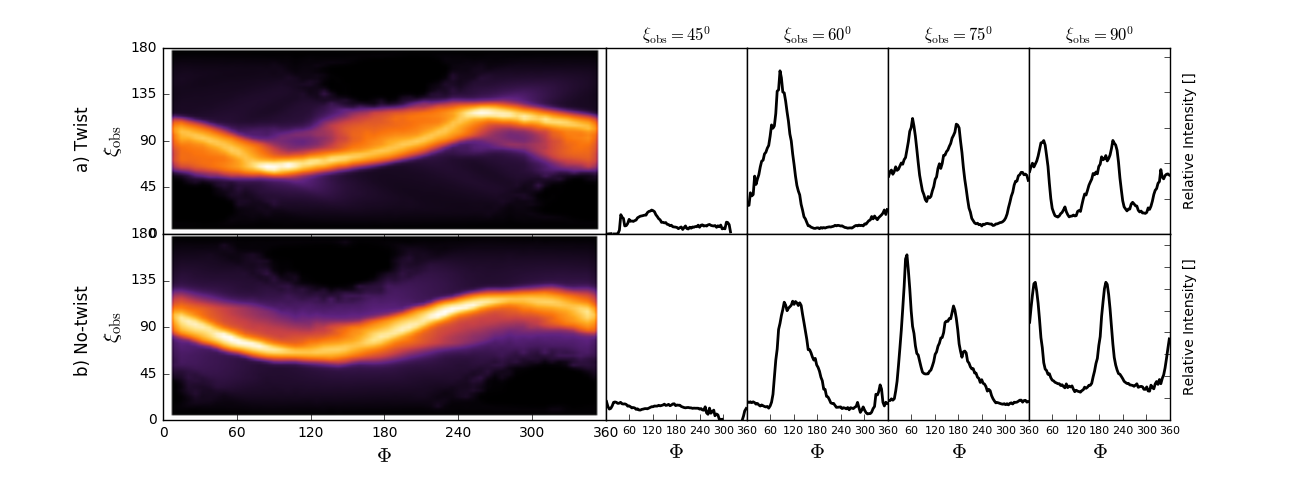}
\end{center}
\caption{\label{fig:Atlas_twist} Effect of the magnetosphere twist on the sky map and light curves for a single neutron star with a FF dipolar field with $\alpha = 30^\circ$ and $r^c_\text{ov} = 0.87$. 
\textit{Upper panel:} Twisted case. 
\textit{Lower panel:} Twist-free case (same as the upper panel of figure~\ref{fig:Atlas}).}
\end{figure}

The effect of such a twist on the sky map can be appreciated in figure~\ref{fig:Atlas_twist}, which displays the sky maps for both the twisted and twist-free cases. 
We find that even for the large current implanted ($\approx10$ times larger than the SL currents), the impact on the light curves is quite modest. This supports the naive expectation that if the twisted bundle of field lines is confined inside the zone of closed field lines of the pulsar, it would not significantly affect its gamma-ray emission in the SL model.\footnote{Further emission could originate at the twisted bundle itself, which by construction would not be captured by the SL model.} The interesting question of whether stable configurations can be achieved for scenarios with the twist footprints crossing (or laying inside) the polar caps, and how they could affect the sky maps, is outside the scope of this work. 

%

%%%%%%%%%%%%%%%%%%%%%%%%%%%%%%%%%%%%%%%%%%%%%%%
%%%%%%%%%%%%%%%%%%%%%%%%%%%%%%%%%%%%%%%%%%%%%%%
\subsection{Offset dipole magnetosphere}
\label{sec:Offset}
%%%%%%%%%%%%%%%%%%%%%%%%%%%%%%%%%%%%%%%%%%%%%%%
%%%%%%%%%%%%%%%%%%%%%%%%%%%%%%%%%%%%%%%%%%%%%%%

Our understanding of stellar collapse offers no particular reason why a star's dipole should be centered after the collapse of its progenitor. The emission characteristics from offset dipoles have been studied before (e.g.~\cite{petri2016radiation, barnard2016effect}). In fact, the picture of NS hot spots emerging recently from NICER suggests not only off-centered fields, but also a field configuration more complex than a dipole~\cite{Bogdanov_2019,Bilous_2019}.
 
It has been recognized that a shift from the stellar center could have strong implications for EM emission from the pulsar in several bands \cite{petri2016radiation, barnard2016effect}. In particular, such off-centered dipolar configurations have been proposed to explain lags between x-ray and radio profiles in ref.~\cite{petri2020joint}.
Additionally, for BHNS binary systems, the field structure would share characteristics with that of a star with an off-centered dipole.

Thus, in this section we consider the magnetic configuration of a dipole displaced from the stellar center either along or perpendicular to the rotational axis. We shift the dipole center, $\boldsymbol{r_o}$, (imposed via boundary conditions at the stellar surface) by
\begin{equation}
\boldsymbol{r_o} (t) = d \left( \sin \delta \, \cos(\Omega t), \sin \delta \, \sin(\Omega t) , \cos \delta \right) 
\label{eq:offset}
\end{equation}
where $d$ is the distance from the stellar center and $\delta$ is the
colatitude. Thus, the relative displacement $d/R$, with respect to the NS
radius $R$, is a relevant parameter. We explore cases with $d/R = 0.6$ and $\delta = \{0, \pi/2 \}$.
These two colatitudes correspond to an offset along and perpendicular to the rotational axis at $t=0$, respectively, specifically along the $z$- and $x$-axis.
We set the rotation rate such that $R/\rho_{\rm LC}=0.25$.

\begin{figure}[!ht]
\begin{center}
\includegraphics[scale=0.2]{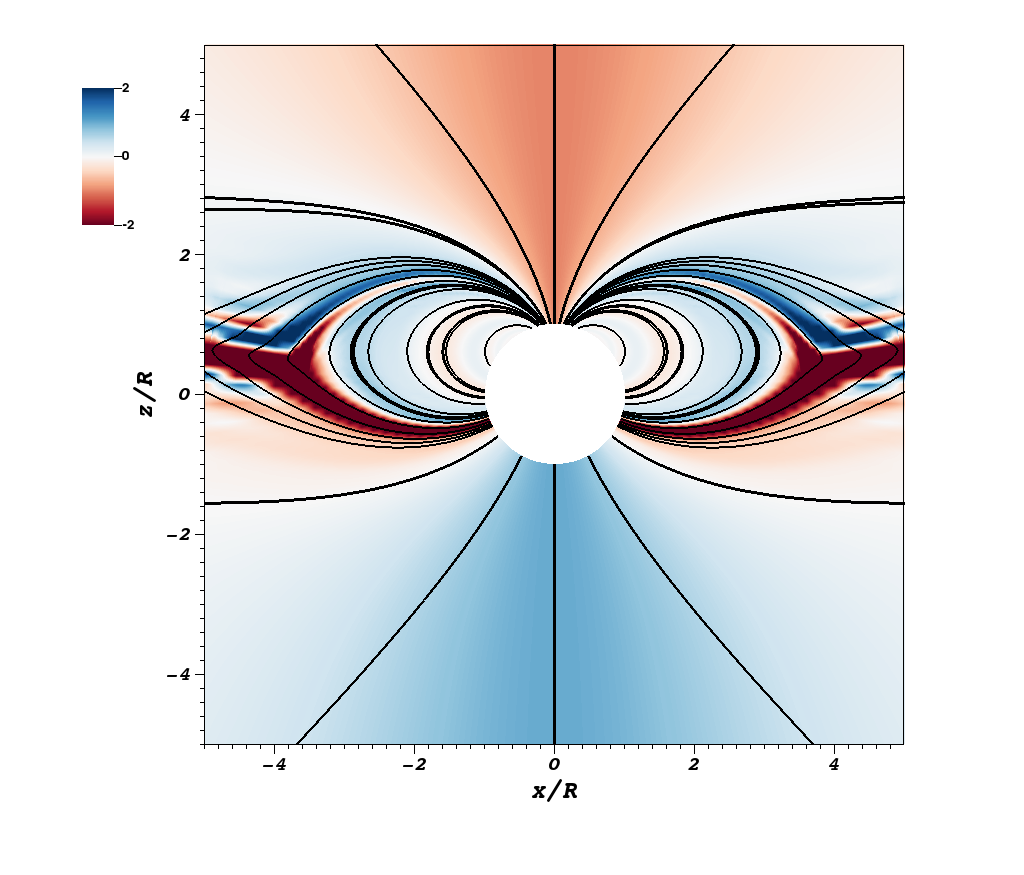}
\includegraphics[scale=0.2]{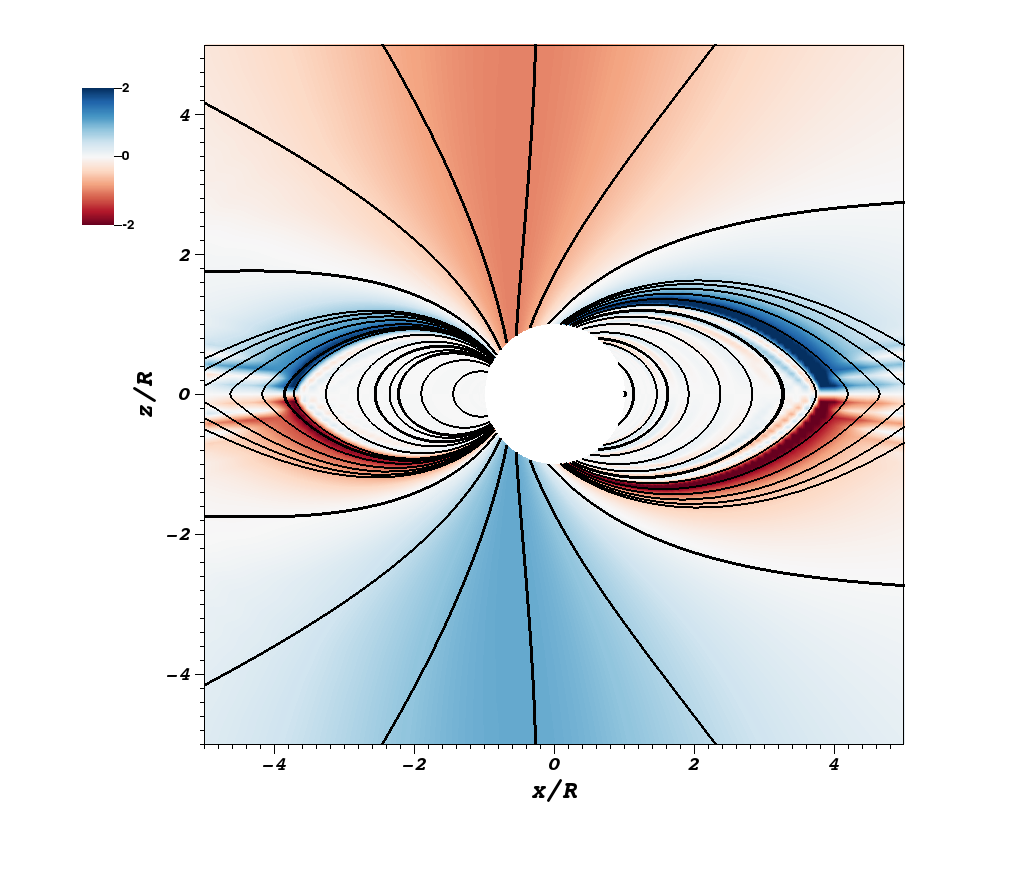}
\end{center}
\caption{Offset dipole solutions after 2.5 periods for parameters $d/R = 0.6$ and $\delta = \{0, \pi/2 \} $ (left and right panels, respectively) as described in eq.~\ref{eq:offset}. The plots show magnetic field lines and coloured parallel electric currents (normalized with respect to $\Omega B / 2\pi$) in the $y=0$ plane. }
\label{fig:offset_single}
\end{figure}

\begin{figure}[h!]
	\centering
    \hfil
    \includegraphics[width=4cm]{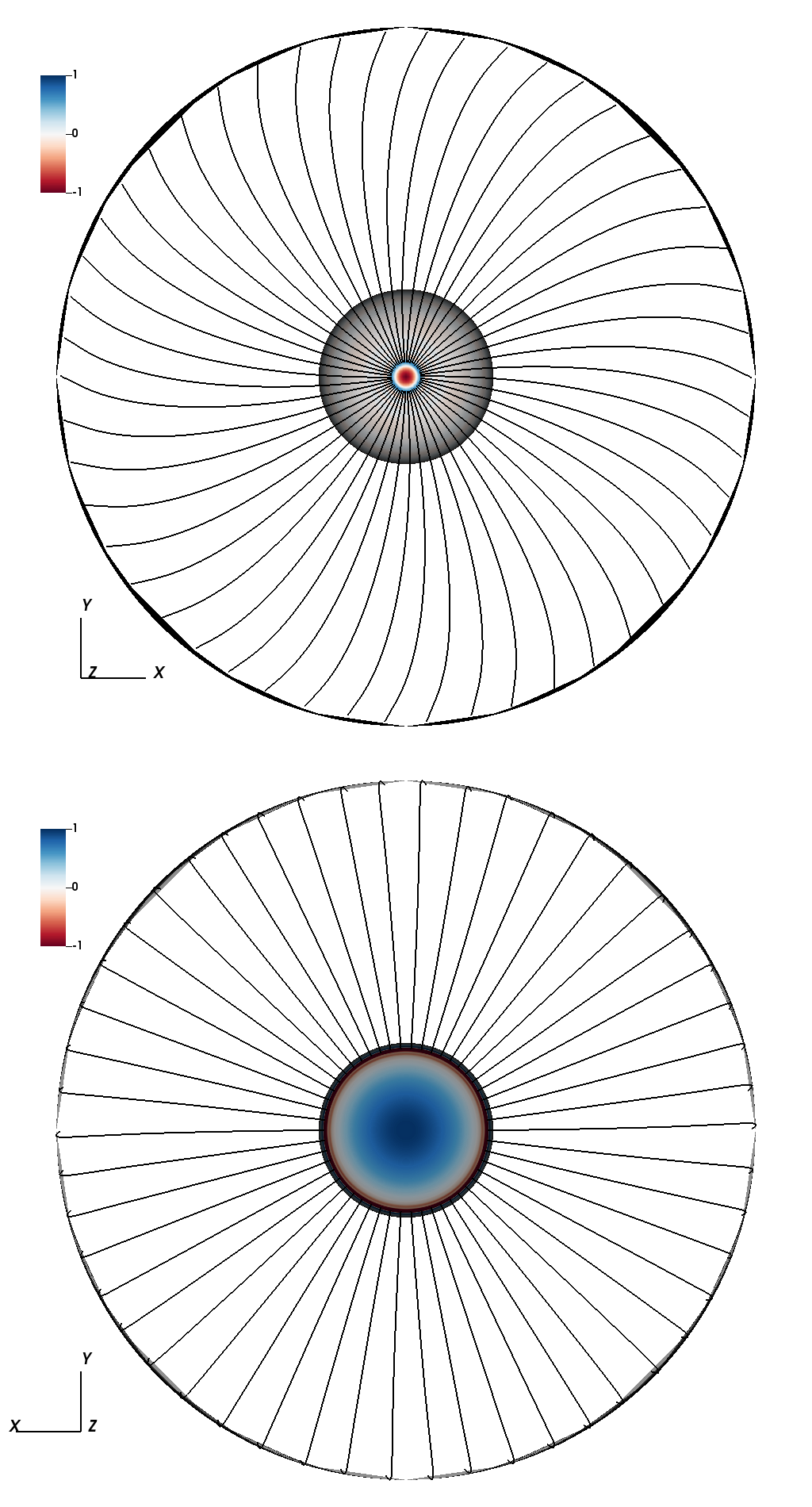}
    \hfil
    \includegraphics[width=8cm]{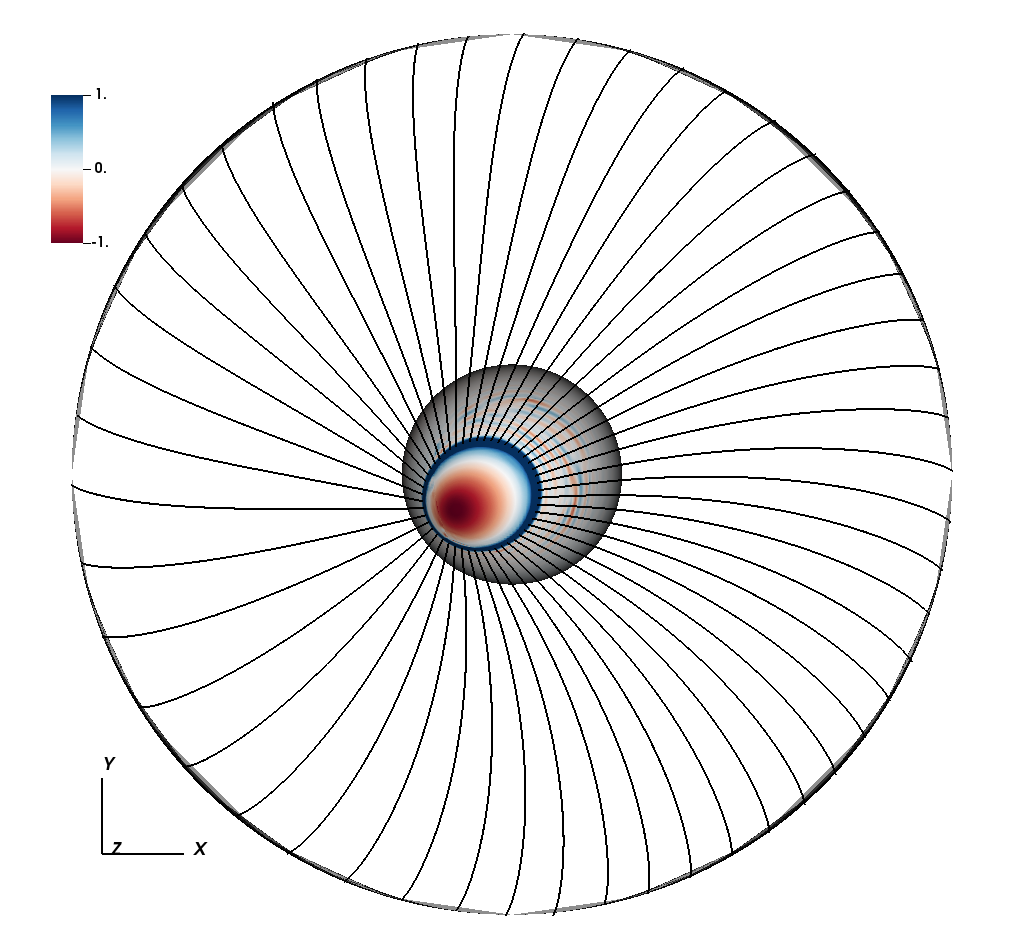}
\caption{LOFLs of single offset dipoles with $d/R = 0.6$ for a $z$-offset (left panel) and an $x$-offset (right panel) after $\sim2.5$ stellar rotations. These are the same solutions shown in figure~\ref{fig:offset_single}. Polar views of the caps are shown with emerging magnetic field lines and coloured parallel electric currents at the NS surface.
The enclosing circles depict the LC, located at $\rho_\text{LC} = 4 R$.
}
\label{fig:offset_single_caps}
\end{figure}

As shown in figure~\ref{fig:offset_single}, the equatorial reflection symmetry of a centered, aligned (i.e. $\alpha = 0^\circ$) dipole is broken by the $z$-offset (left panel), whereas the azimuthal rotation symmetry is broken by the $x$-offset (right panel). The LOFLs and polar caps after $\sim2.5$ stellar rotations are viewed from polar vantage points in
figure~\ref{fig:offset_single_caps}.

\begin{figure}[!ht]
\begin{center}
\includegraphics[width=17cm]{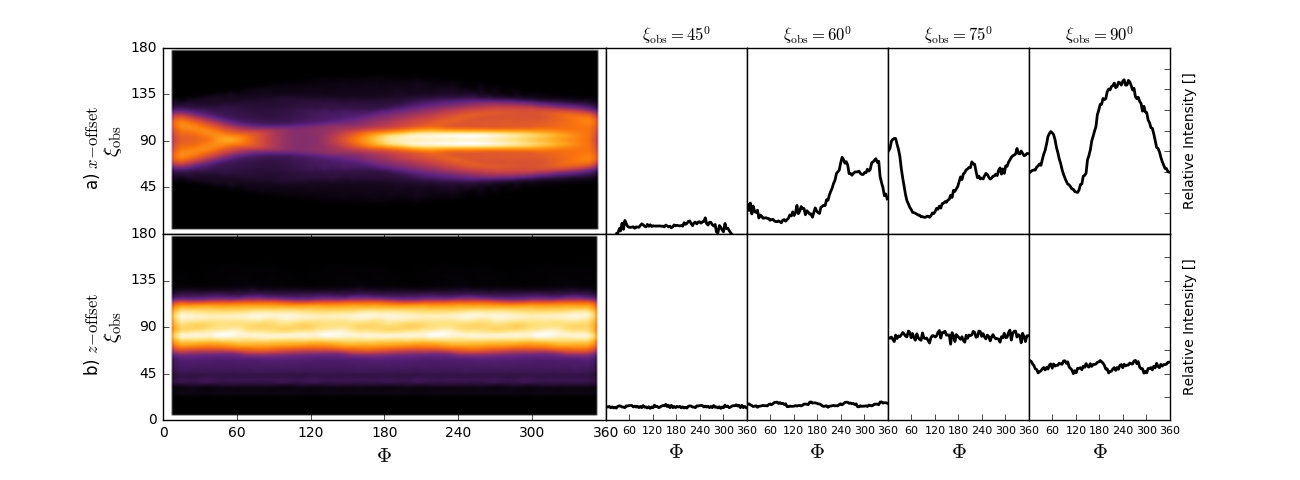}
\end{center}
\caption{\label{fig:atlas_offset_aligned} Sky maps and light curves for the FF aligned ($\alpha=0^\circ$) offset dipole magnetospheres of the single NSs shown in figure~\ref{fig:offset_single}. In both cases, the offset is $d/R = 0.6$, and the central open volume coordinate is $r^c_\text{ov} = 0.87$.}
\end{figure}

Accordingly, figure~\ref{fig:atlas_offset_aligned} shows the resulting sky maps and light curves for each of these configurations. As expected, the symmetry of the emission intensity is broken according to the direction of the offset. As described in section~\ref{sec:pure}, the strength of emission is proportional to the length of particle trajectories. Since particles follow field lines, the strength of emission is proportional to the length of those field lines until the imposed cutoff radius. We see this effect in the sky map for the $z$-offset case in panel $(b)$ of figure~\ref{fig:atlas_offset_aligned}: the northern hemisphere ($\xi_{\rm obs}<90^\circ$) is brighter than the southern hemisphere. Similarly, the direction opposite to the $x$-offset is brighter than the direction of the $x$-offset (see panel $(a)$ of figure~\ref{fig:atlas_offset_aligned}). Note also that the $z$-offset is axisymmetric, so fluctuations in the light curves as a function of phase $\Phi$ are numerical in nature.

\begin{figure}[!ht]
\begin{center}
\includegraphics[width=17cm]{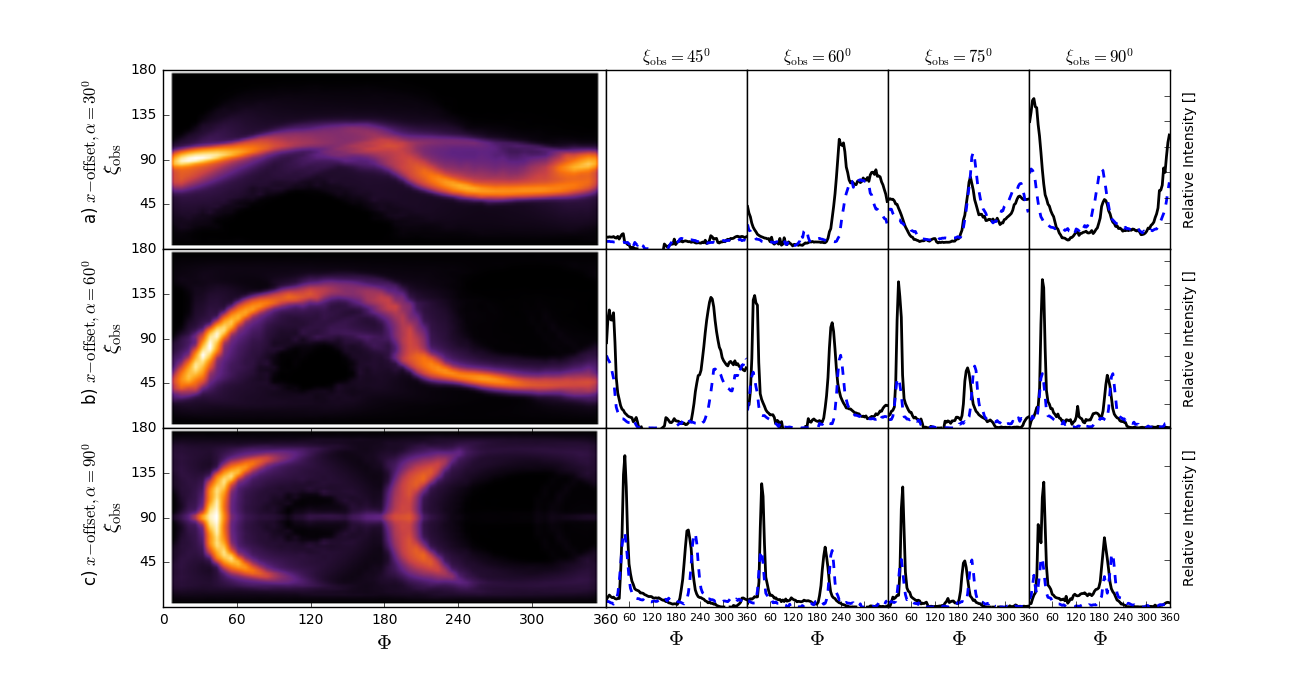}
\end{center}
\caption{\label{fig:atlas_offset_inclined-x} Sky maps and light curves for FF inclined $x$-offset dipole magnetospheres in a single NS. In all cases, the offset is $d/R = 0.6$. Notice the intensity asymmetry across $\Phi=180^\circ$. The center of open volume coordinates $r^c_\text{ov}$ for each inclination angle $\alpha$ is chosen as in figure~\ref{fig:Atlas}. For the ease of comparison with light curves from the centered dipole solution, we have superimposed the light curves from figure~\ref{fig:Atlas} using dashed-blue lines.}
\end{figure}

\begin{figure}[!ht]
\begin{center}
\includegraphics[width=16.9cm]{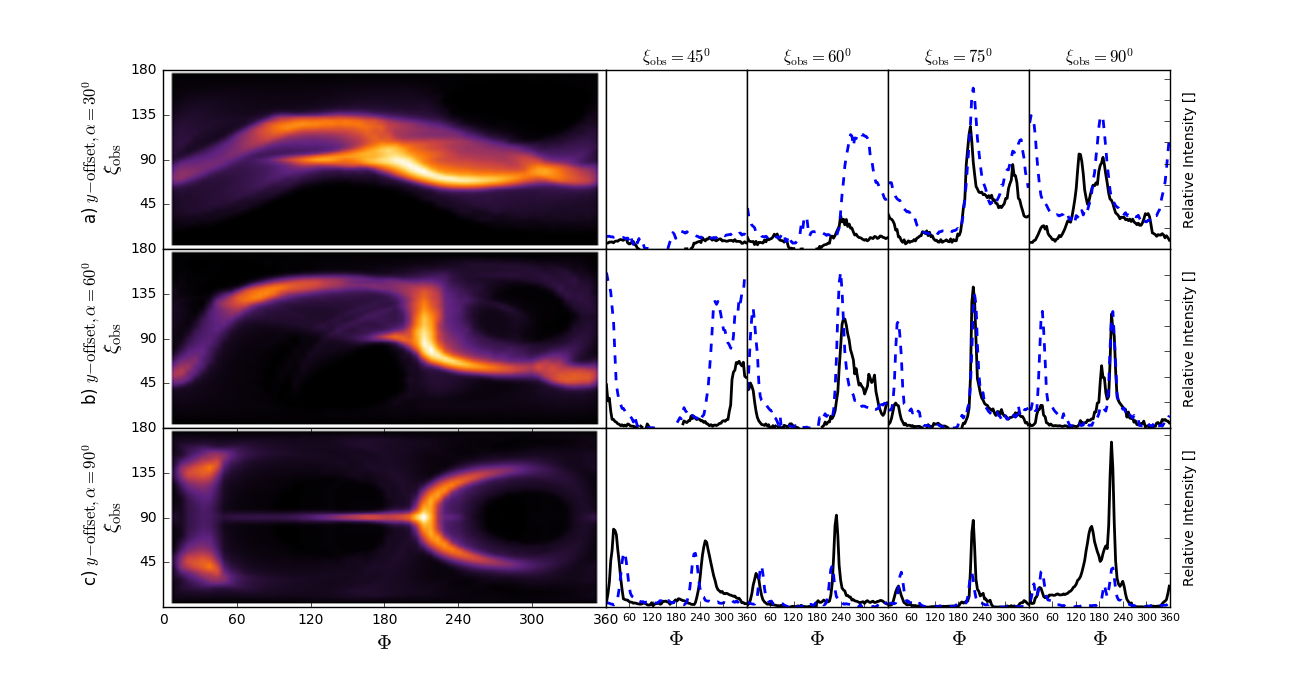}
\end{center}
\caption{\label{fig:atlas_offset_inclined-y} Sky maps and light curves for FF inclined $y$-offset dipole magnetospheres in a single NS. In all cases, the offset is $d/R = 0.6$. The center of open volume coordinates $r^c_\text{ov}$ for each inclination angle $\alpha$ is chosen as in figure~\ref{fig:Atlas}. For comparison with light curves from the centered dipole, we have superimposed the light curves from figure~\ref{fig:Atlas} using dashed-blue lines.}
\end{figure}

\begin{figure}[!ht]
\begin{center}
\includegraphics[width=16.9cm]{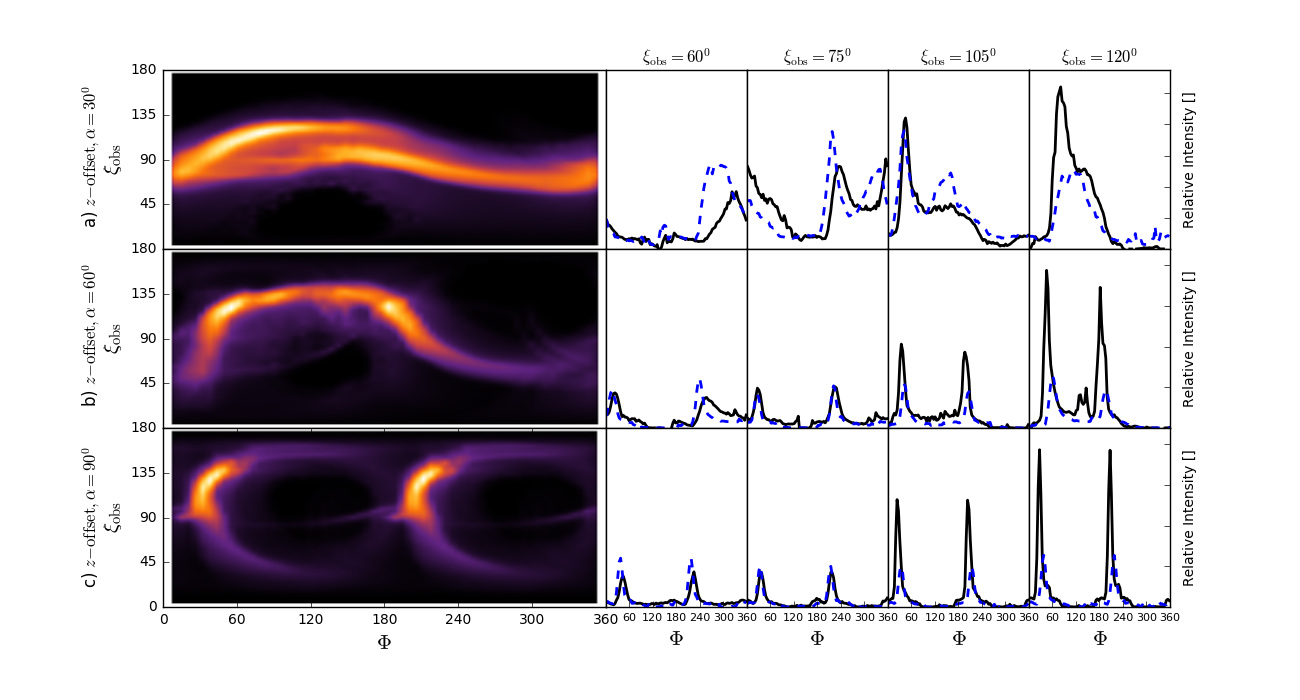}
\end{center}
\caption{\label{fig:atlas_offset_inclined-z} Sky maps and light curves for FF inclined $z$-offset dipole magnetospheres in a single NS. In all cases, the offset is $d/R = 0.6$. 
Notice the intensity asymmetry across the equator, $\xi_\text{obs} = 90^\circ$.
The center of open volume coordinates $r^c_\text{ov}$ for each inclination angle $\alpha$ is chosen as in figure~\ref{fig:Atlas}. Notice that light curves are taken at observation angles different from other cases. For comparison, we have superimposed light curves from the centered dipole solution using dashed-blue lines.}
\end{figure}

Next, we allow an inclination (i.e. non-zero $\alpha$) within the $x$-$z$ plane to the offset dipole.
Figures~\ref{fig:atlas_offset_inclined-x}, \ref{fig:atlas_offset_inclined-y}, and \ref{fig:atlas_offset_inclined-z} present both the sky maps and representative light curves for the $x$-, $y$-, and $z$-offsets, respectively.\footnote{Note that a magnetic inclination in the $x-z$ plane makes $x$- and $y$-offsets distinct from each other.}
Comparing these results with the centered case shown in figure~\ref{fig:Atlas},
one can see clear intensity asymmetries.
For the $x$-offset shown in figure~\ref{fig:atlas_offset_inclined-x}, the intensity is asymmetric
in the azimuthal viewing angle (across $\Phi=180^\circ$).
For the $z$-offset shown in figure~\ref{fig:atlas_offset_inclined-z}, the intensity is asymmetric
across the equator ($\xi_\text{obs} = 90^\circ$). Whereas for the $y$-offset in figure~\ref{fig:atlas_offset_inclined-y}, since the inclination is in the $x$-$z$ plane, both symmetries are generally broken, yielding larger departures from figure~\ref{fig:Atlas}.   

Although the generic double peaks of light curves in figure~\ref{fig:Atlas} are preserved in
figures~\ref{fig:atlas_offset_inclined-x},~\ref{fig:atlas_offset_inclined-y},~and~\ref{fig:atlas_offset_inclined-z},
a remarkable consequence of the $x$-offset is a
more pronounced height difference between peaks, which is clearer as the inclination angle increases. 
Beyond single star systems with offset dipoles,  such a qualitative behavior would also be expected in a non-vacuum binary system where 
only one star is strongly magnetized. The observed impact on peak height can thus help characterize
the system. On the other hand, the $z$-offset does not produce large changes in the shapes of light curves, but primarily produces differences in 
overall intensity across observers. Thus, from the observational point of view, polar offsets are more difficult to detect 
than equatorial offsets.  Of course, the ability to scrutinize pulsar systems at all depends on detectability prospects, and we recall that the roughly $70$ detected 
gamma ray pulsars lie in our galaxy or quite nearby~\cite{Saz_Parkinson_2010,doi:10.1146/annurev-astro-081913-035948,PSRGAMMA801}.
%

%%%%%%%%%%%%%%%%%%%%%%%%%%%%%%
%%%%%%%%%%%%%%%%%%%%%%%%%%%%%%
\section{Beyond single dipoles: Multipolar structure and binaries}
\label{sec:Binaries}
%%%%%%%%%%%%%%%%%%%%%%%%%%%%%%
%%%%%%%%%%%%%%%%%%%%%%%%%%%%%%
We now turn our attention to high-energy emission from sources with richer global field structure. As a first step towards describing binaries and motivated in part by our observations above, 
we consider a single spherical surface enclosing a binary as a model of the large-scale magnetic field and CS. 
The single spherical surface is dressed with the individual magnetic moments of the binary components, and we refer to this as the \emph{enclosing surface approximation}. A single star having a superposition of dipoles, for example, is an interesting scenario in its own right given the results from NICER. But such a system can also represent an approximation of a binary with dipole magnetizations. We regard the enclosing surface approximation as a simple tool to rapidly explore binary magnetic field configurations. Systems identified as sufficiently interesting can then be explored in more depth (with full general relativity, for example). We first introduce the enclosing surface approximation for a pulsar on an artificially prescribed circular orbit, which we approximate as a single star with an offset dipole.  We show that the orbiting pulsar yields radiation sky maps consistent with that of a single star enclosing
an offset dipole.
We then focus on the superposition of two offset dipoles---regarded either as a single star with a multipolar magnetic field or a BNS system. Finally, we study the emission associated with a fully nonlinear, numerical evolution of a BHNS binary in general relativity using the methods described in section~\ref{sec:Currents}.

%%%%%%%%%%%%%%%%%%%%%%%%%%%%%%
\subsection{Single stars and the enclosing surface approximation for binaries}
\label{sec:Enclosing}
%%%%%%%%%%%%%%%%%%%%%%%%%%%%%%
Observations such as those of NICER motivate the study of multipolar field configurations in isolated stars. Indeed, recent theoretical
efforts have begun exploring examples even including gravitational effects (see refs.~\cite{Gralla:2016fix,Gralla:2017nbw}).
Another motivation relevant for the work here is that binaries, even those consisting of two simple dipoles,  display an
effectively multipolar field structure in the far zone.

With these motivations, we employ an enclosing surface model in which the field beyond a given radius encompassing the binary is approximated by the binary's constituent magnetic moments enclosed by a spherical, perfectly conducting surface. 
Notice that such a surface can be chosen to lie just outside two
orbiting NSs and will be well inside the orbital LC even when the binary is close to merger. 
With this model, we can either:
(i) represent a binary's large-scale magnetic field behavior with a single effective star that rotates at the binary's orbital frequency,
or (ii) represent a single star that genuinely has two magnetic dipoles.

The model allows for a straightforward extension of the single star analysis, 
producing preliminary radiation sky maps that will help to interpret those from numerically evolved %the actual 
binaries to be presented elsewhere. 
In particular, this model allows for capturing the main
dynamical effects associated with the launching and reflection of Alfv\'en
waves. 
With one~(BHNS) or two~(BNS) properly arranged magnetic dipoles, it
yields approximate solutions for the exterior,
common magnetosphere of the binary system.

As formulated, a limitation of this model is that it can only represent NSs tidally locked in their orbit.
While this condition is not expected to be realized in nature~\cite{1992ApJ...400..175B}, the model nevertheless
captures certain essential features of the far-field configuration and of the high-energy emission as we illustrate below.

%%%%%%%%%%%%%%%%%%%%%%%%%%%%%%%%%%%
\subsubsection{Single dipole scenario}
\label{sec:Enclosing_1dipole}
%%%%%%%%%%%%%%%%%%%%%%%%%%%%%%%%%%%

To illustrate the value of this approach,  we study a single offset dipole within the
enclosing surface approximation. The far-field behavior of this system models that of a 
compact object binary with only one magnetized component. We compare the resulting sky map with
that resulting from the evolution of a magnetized NS on an artificially prescribed circular orbit, as presented in ref.~\cite{Orbiting}.
This latter evolution seeks to mimic the trajectory of a star tidally locked within a compact object binary. As such, the stellar spin $\Omega_*$ equals the orbital frequency $\Omega_\text{o}$.
For a consistent comparison, the center of the enclosing sphere matches the 
orbital center of the orbiting NS, and other parameters are chosen accordingly. The LC radius $\rho_{\rm LC}$ compares with the enclosing surface radius $\mathcal{R}$ as $\mathcal{R}/\rho_{\rm LC}=0.25$

\begin{figure}[!ht]
\begin{center}
\includegraphics[scale=0.2]{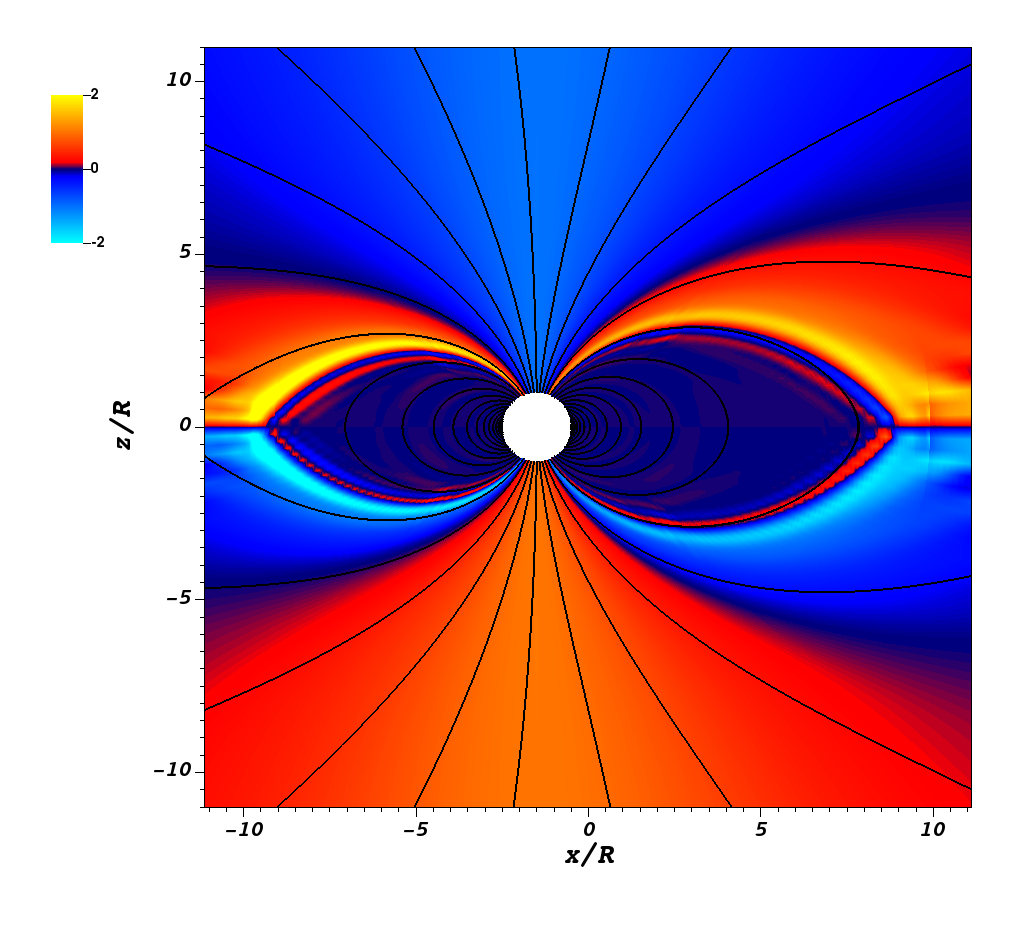}
\hspace{0.7cm}
\includegraphics[scale=0.2]{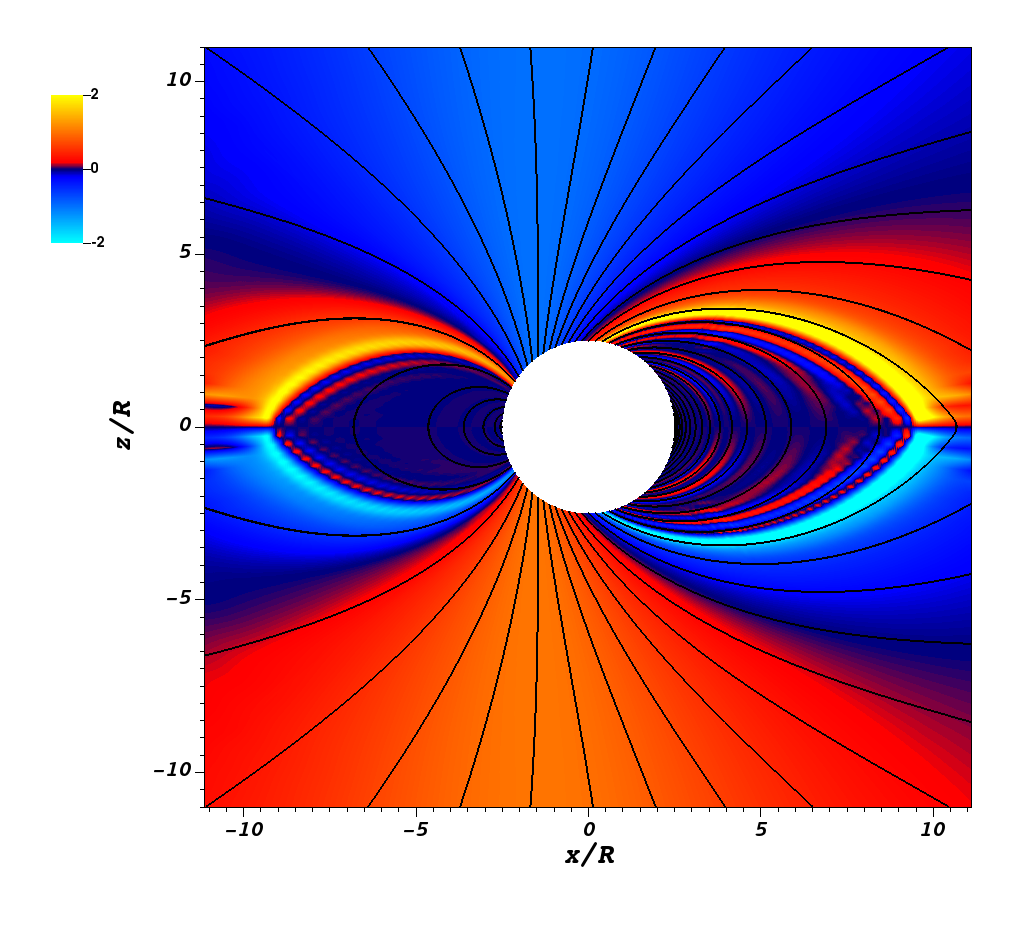}
\end{center}
\caption{ 
Comparison of an orbiting NS (\textit{left panel}) with
the enclosing surface approximation of an offset dipole (\textit{right panel}).
The orbiting NS simulation was presented in ref.~\cite{Orbiting}. It has
orbital separation $a=3R$ (where $R$ is the NS radius) after $\sim 2.5$ orbits, and its
orbital frequency $\Omega_\text{o}$ matches the star's angular velocity $\Omega_*$ with value $\Omega_\text{o}=\Omega_*= 0.1 c/R$.
The approximation on the right panel includes an offset dipole inside
a rotating surface (white circle) of radius $\mathcal{R}=2.5R$ with parameters chosen to match the evolution on the left. Note that the LC radius compares with $\mathcal{R}$ as $\mathcal{R}/\rho_{\rm LC}=0.25$.
Magnetic field lines (black lines) and coloured parallel electric currents (normalized with respect to $\Omega B/2\pi$) are shown on the $y=0$ plane,
and their significant agreement, particularly in the far field, suggests that
the model captures the large-scale features of the magnetosphere as discussed in section~\ref{sec:Enclosing_1dipole}.
\label{fig:test_fat-star}}
\end{figure}

Figure~\ref{fig:test_fat-star} shows the resulting magnetic field structure from both solutions. The left panel shows the spherical surface of the star in white, orbiting about the origin. The right panel shows a larger spherical surface centered on the origin that encloses a dipole offset to the left.\footnote{The offset dipole presented here is in fact the same solution as shown in the right panel of figure~\ref{fig:offset_single} from section~\ref{sec:Offset}, suitably scaled for this comparison.} The parallel electric currents and field structure are strikingly similar, particularly in the far field. 

\begin{figure}[!ht]
\begin{center}
\includegraphics[width=17cm]{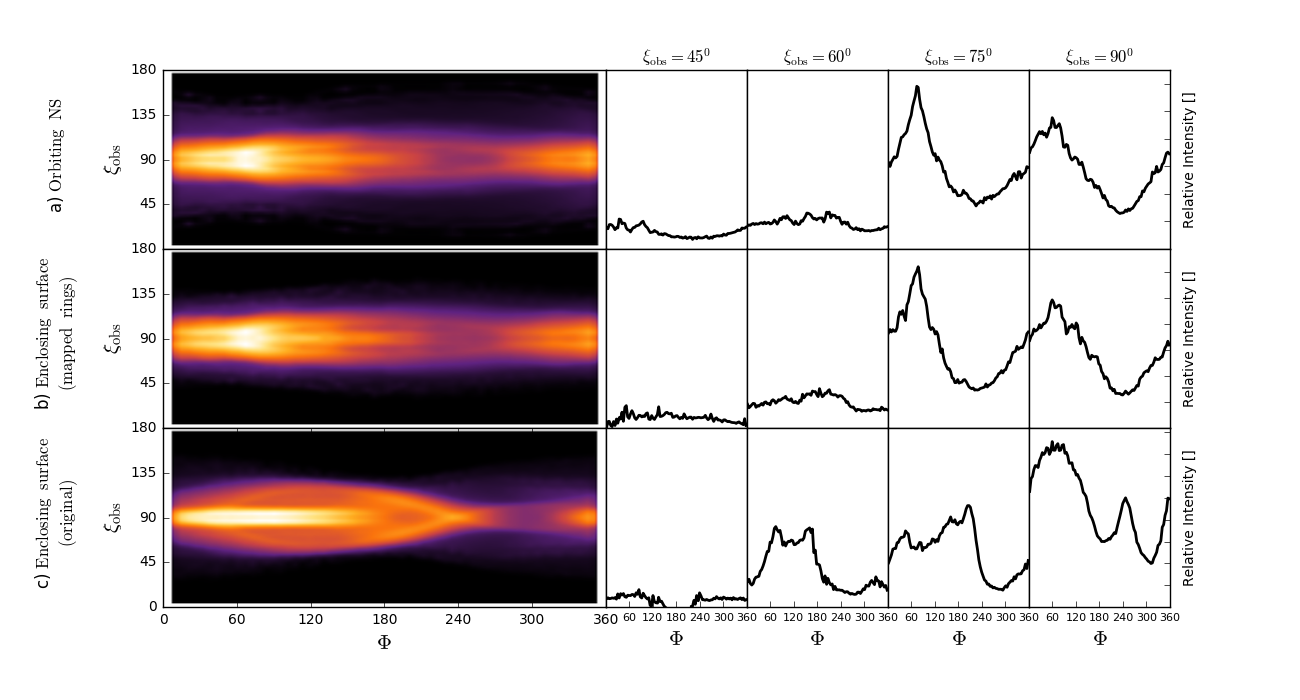}
\end{center}
\caption{\label{fig:atlas_oNS} Comparison of the sky maps and light curves from the
enclosing surface approximation (\textit{middle and bottom panels}) and the evolution of the
orbiting NS (\textit{top panel}) from the right panel of figure~\ref{fig:test_fat-star}.
The qualitative agreement is discussed in section~\ref{sec:Enclosing_1dipole}. In the \textit{middle panel}, 
emission rings on the enclosing surface are obtained by mapping emission rings ($r_\text{ov} < 1$) from the orbiting NS surface to the enclosing surface via magnetic field lines.
The \textit{bottom panel} results from emission rings determined on the enclosing surface of the right panel of figure~\ref{fig:test_fat-star} in the usual way.
It is indeed a reproduction of the top panel of figure~\ref{fig:atlas_offset_aligned}, except shifted in $\Phi$ for ease of comparison. In all cases, $r_\text{ov}^\text{c} = 0.87$.
}
\end{figure}

Given this agreement, it is not surprising that the resulting sky maps displayed in figure~\ref{fig:atlas_oNS} are similar as well. The sky map corresponding to the x-offset dipole
(bottom panel of figure~\ref{fig:atlas_oNS})
is produced by photon emission along lines emerging from a set of rings inside the polar cap on the enclosing surface. This polar cap is determined by finding field lines that are tangent to the LC.
In the SL model (see section~\ref{sec:pure} for a detailed description), the construction of emission rings with open volume coordinate $r_\text{ov} < 1$ inside this cap yields a set of rings that is independent of the set of rings obtained using the same procedure on the orbiting NS surface.
Thus, even though the caps themselves are consistent among the two FF solutions, their derived emission regions are not necessarily equivalent. A direct application of the SL model to the enclosing surface solution yields the sky map in
the lower panel of figure~\ref{fig:atlas_oNS}, which is similar to, but more distinct than, the one produced by the orbiting NS (top panel of figure~\ref{fig:atlas_oNS}). 
On the other hand, any alternative models which assume emission only along the LOFLs would be well-captured by the caps on the enclosing surface.
To illustrate how well the magnetic field is approximated in the enclosing surface system, we mapped emission rings from the orbiting NS surface to the enclosing surface along magnetic field lines. For this new set of emitting rings on the enclosing surface, the resulting sky map is displayed in the middle panel of figure~\ref{fig:atlas_oNS}, and is essentially identical to the one produced by the orbiting NS in the top panel of the same figure.
Of course, when applying the enclosing surface approximation in practice, one does not know the geometry of the objects being enclosed. It may be possible to improve the cap-shrinking procedure, such that it takes into account geometrical information about the orbiting stars. We leave such a potential improvement for future work.

%
%%%%%%%%%%%%%%%%%%%%%%%%%%%%%%%%%%%
\subsubsection{Double dipole scenario}
\label{sec:Enclosing_2dipoles}
%%%%%%%%%%%%%%%%%%%%%%%%%%%%%%%%%%%
The double dipole initialization superposes two dipole magnetic fields
with two different dipole moments, $\boldsymbol{m}$ and $\boldsymbol{m}^\prime$,
which are symmetrically displaced from the center.
Such a configuration could represent a simple model of the far field of a binary system
with both constituents magnetized, or it could model a single neutron star with
a magnetic field configuration quite distinct from a single centered dipole (for other configurations
examined recently see~\cite{Gralla:2017nbw}).
Letting the initial offset be $\pm x_0 \hat{\boldsymbol{x}}$, the double dipole field is given by
\begin{eqnarray}\label{sup1}
\boldsymbol{B}_{\mathrm{DD}} &=& \boldsymbol{B}(\boldsymbol{m},\boldsymbol{r}-x_0 \hat{\boldsymbol{x}}) +
\boldsymbol{B}(\boldsymbol{m}^\prime,\boldsymbol{r}+x_0 \hat{\boldsymbol{x}}) \nonumber \\
&=& \frac{\mu_0}{4\pi} \left[ 3 (\boldsymbol{r} - x_0 \hat{\boldsymbol{x}}) \frac{\boldsymbol{m}\cdot (\boldsymbol{r}-x_o\hat{\boldsymbol{x}})}{(r^2-2xx_0+x_0^2)^{5/2}} - \frac{\boldsymbol{m}}{(r^2-2xx_0+x_0^2)^{3/2}}\right] \nonumber\\
&+&\frac{\mu_0}{4\pi} \left[ 3 (\boldsymbol{r} + x_0 \hat{\boldsymbol{x}}) \frac{\boldsymbol{m}^\prime\cdot (\boldsymbol{r}+x_o\hat{\boldsymbol{x}})}{(r^2+2xx_0+x_0^2)^{5/2}} - \frac{\boldsymbol{m}^\prime}{(r^2+2xx_0+x_0^2)^{3/2}}\right]. 
\end{eqnarray}
We focus on two particular cases, which we call ``up/up'' ($\boldsymbol{m} = \boldsymbol{m}^\prime = m \hat{\boldsymbol{z}}$) and ``up/down'' ($\boldsymbol{m}=m\hat{\boldsymbol{z}}$ and $\boldsymbol{m}^\prime=-m\hat{\boldsymbol{z}}$).

Employing the enclosing surface approximation with a sphere of radius $\mathcal{R}$,
we set the offset $x_0 =  0.6 \mathcal{R}$. 
In our numerical simulations, we set the angular velocity of the system $\Omega = 0.25 c/\mathcal{R}$, and thus the LC radius 
is $\rho_\text{LC} = c/\Omega = 4 \mathcal{R}$.
We produce sky maps using snapshots of the FF field from the up/up and up/down configurations once they have reached a stationary state,
typically after $2.5$ rotation periods.

\begin{figure}[!ht]
\begin{center}
\includegraphics[scale=0.2]{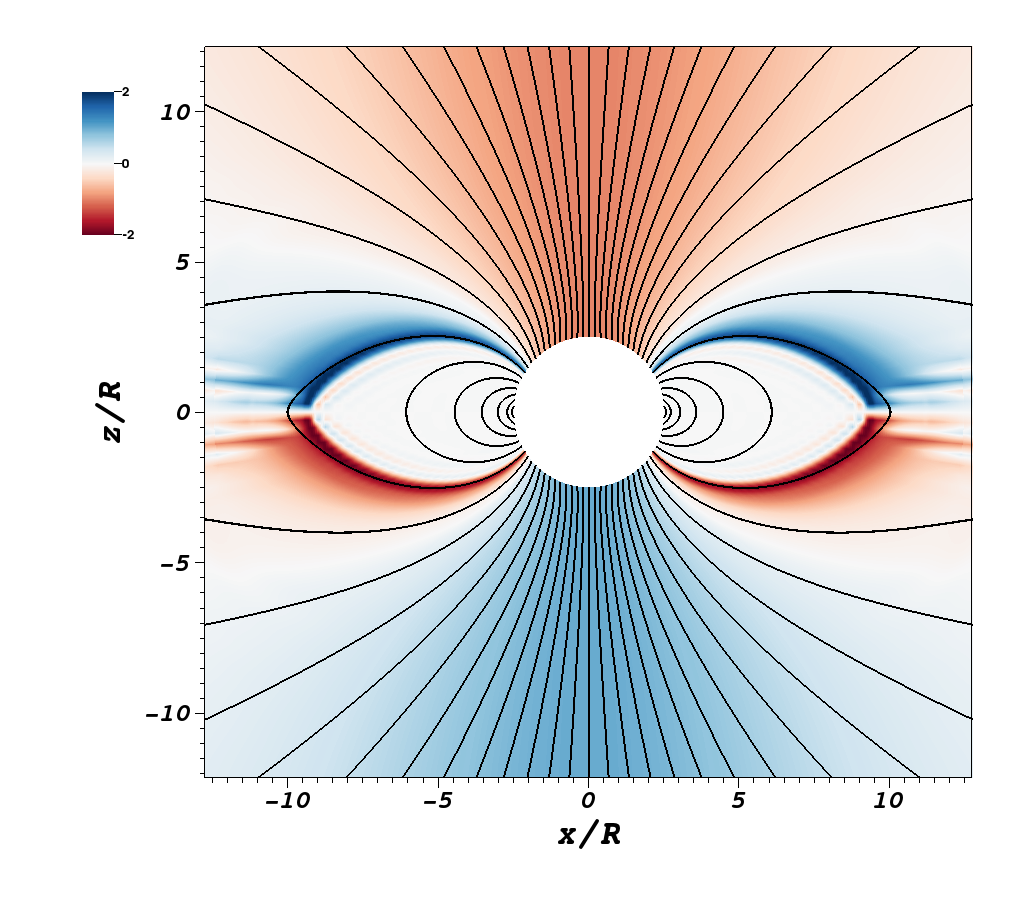}
\includegraphics[scale=0.2]{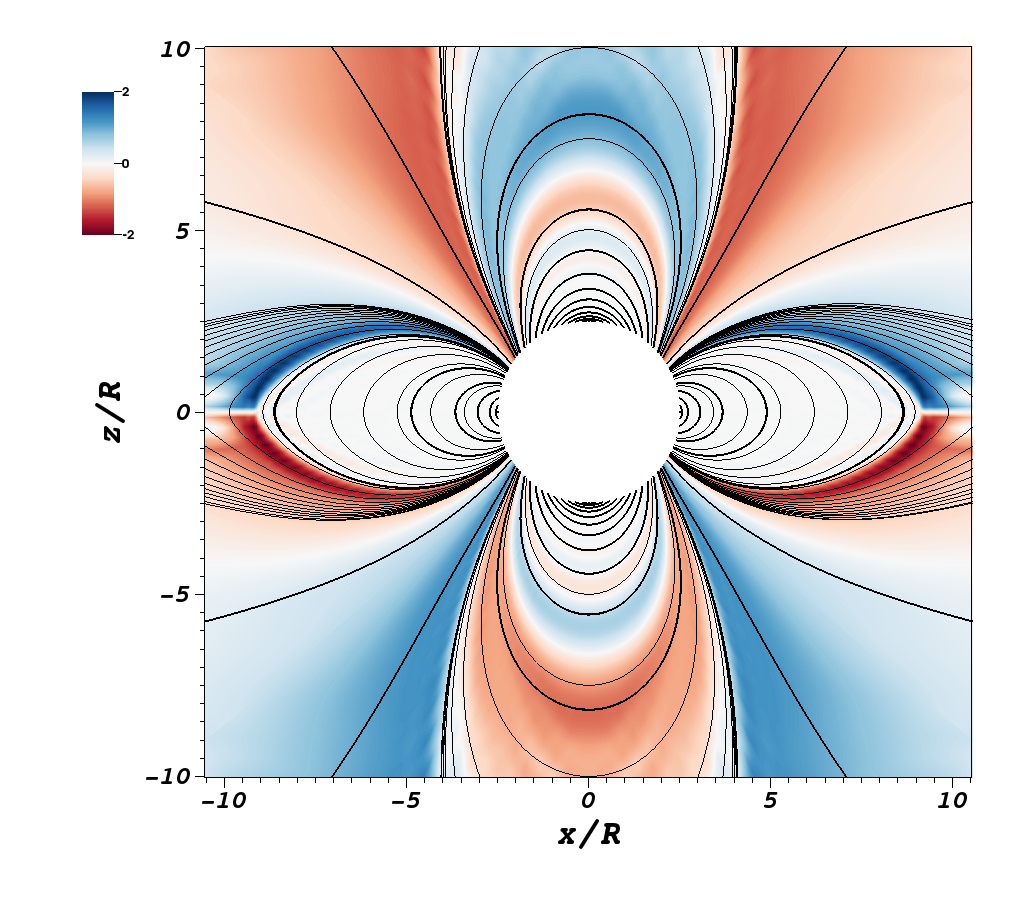}
\end{center}
\caption{Double dipole configurations inside the enclosing surface model with sphere radius $\mathcal{R}$. 
Magnetic field lines and parallel electric currents (normalized with respect to $\Omega B / 2\pi$) on the $y=0$
plane are shown for the up/up (\textit{left panel}) and up/down (\textit{right panel}) configurations. The configurations obtained
are similar to those obtained in evolutions of binary neutron stars within full general relativity 
employing relativistic, resistive magnetohydrodynamics presented in ref.~\cite{ponce2014interaction}.
}
\label{fig:double_dipole}
\end{figure}

\begin{figure}[!ht]
\begin{center}
\includegraphics[scale=0.15]{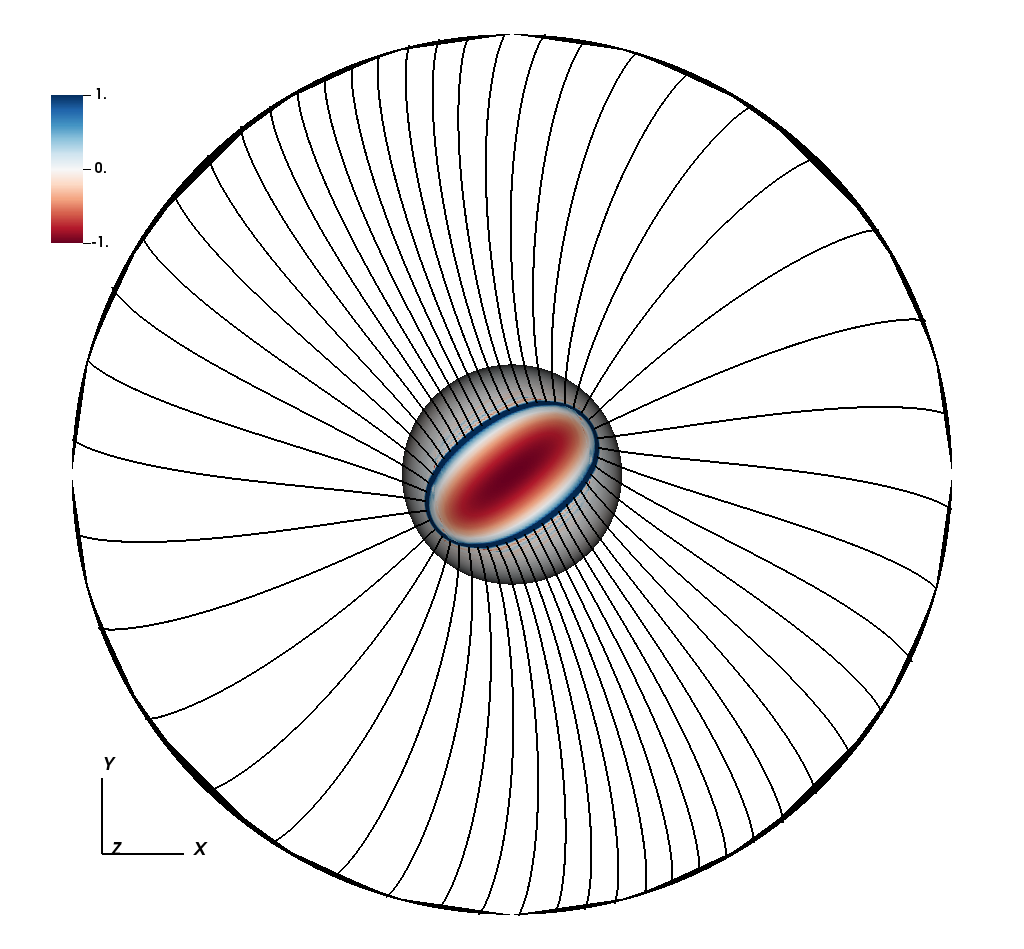}
\includegraphics[scale=0.15]{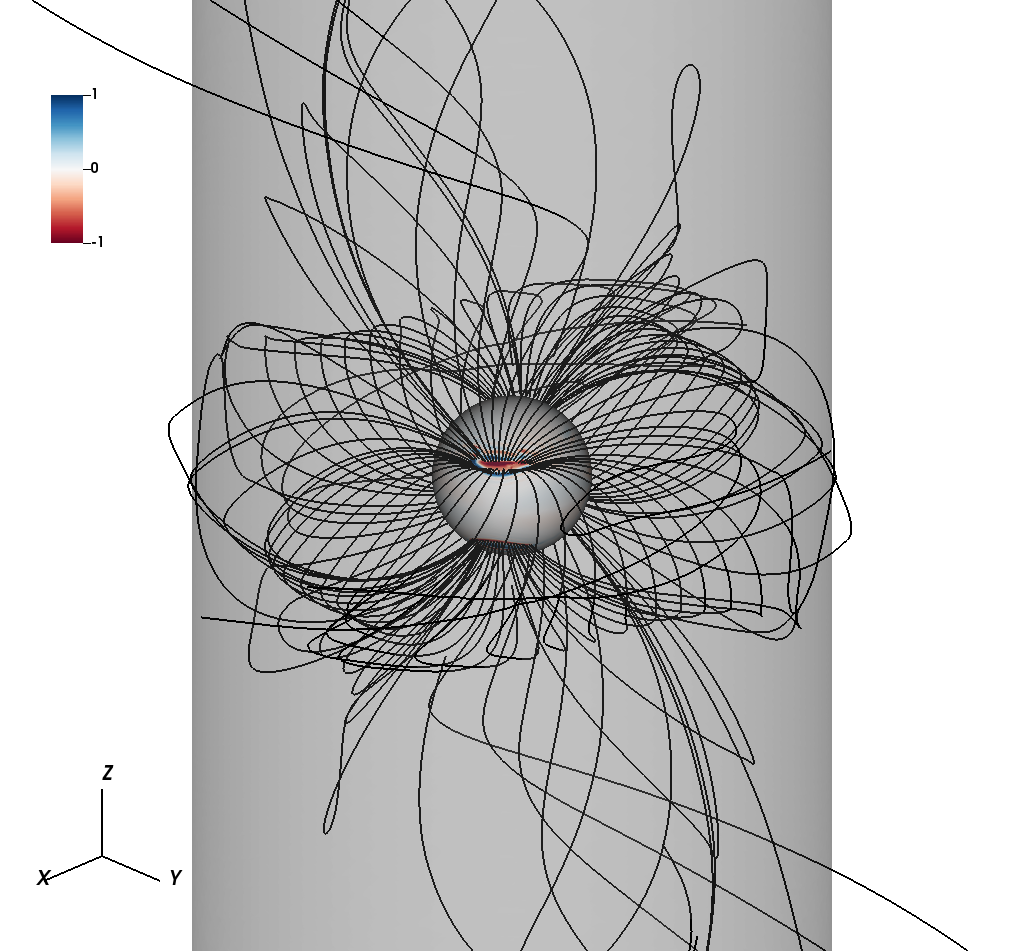}
\end{center}
\caption{LOFLs and parallel electric currents (normalized with respect to $\Omega B / 2\pi$) coloured at the stellar surface, for the up/up (\textit{left panel}) and up/down (\textit{right panel}) double dipole configurations.
The up/up case has two magnetic caps of elliptic shape. The up/down configuration has four caps (more clearly seen in the right panel of figure~\ref{fig:double_dipole}) though only one is plainly
visible here.
}
\label{fig:Lines_2dip_uu}
\end{figure}

Figure~\ref{fig:double_dipole} shows the magnetic field lines and parallel electric currents on
the $y=0$ plane resulting from the up/up (left panel) and up/down (right panel) configurations.
Figure~\ref{fig:Lines_2dip_uu} instead shows the LOFLs emanating from the polar cap boundaries of these configurations. The up/up case is shown in the left panel from a polar vantage point (3D coordinate axes shown), whereas the up/down case is shown from an oblique vantage point in the right panel. The LC is the large circle (left panel) or gray cylinder (right panel).
The field configuration in the up/up case resembles
the single, aligned dipole solution. However, the broken axial symmetry in the magnetic field is quite apparent from the polar vantage point; the polar cap is ellipsoidal and the LOFLs are more dense in the direction orthogonal to the dipole offsets.
When viewed in the plane of the dipole moments (left panel of figure~\ref{fig:double_dipole}), the
field lines resemble a single dipole. 

In contrast to the up/up case, the field topology of the up/down configuration is much more involved. The field has a quadrupolar character, as illustrated in the right panel of figure~\ref{fig:double_dipole}. This complexity gives rise to four different caps, two per dipole---see the right panel of figure~\ref{fig:Lines_2dip_uu}, which shows one of these magnetic caps from an oblique vantage point.

\begin{figure}[!ht]
\begin{center}
\includegraphics[width=17cm]{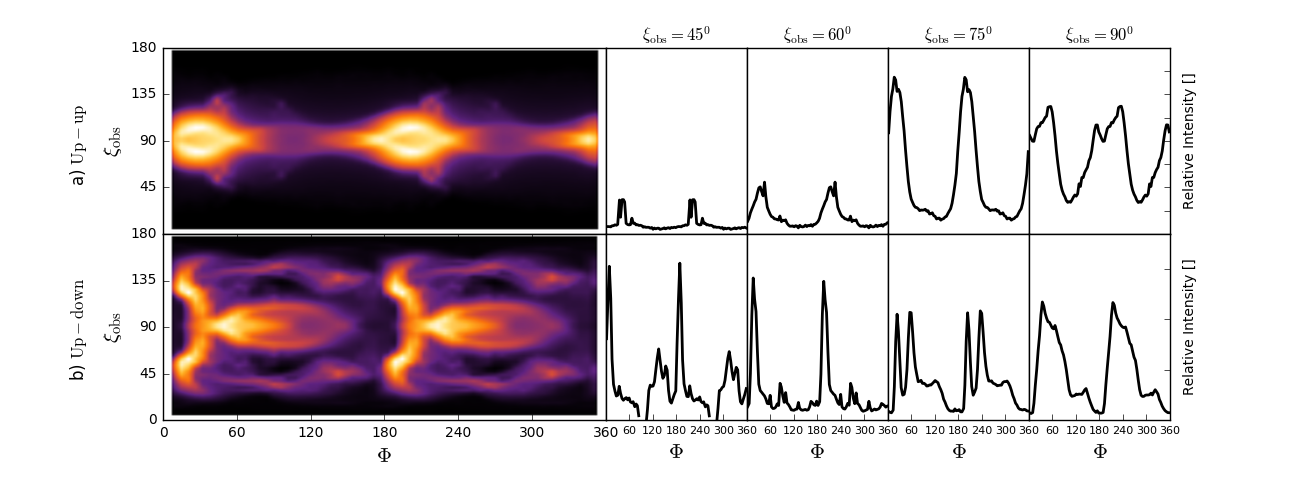}
\end{center}
\caption{\label{fig:atlas_double_dipole} Sky maps and light curves for the two FF, double dipole magnetospheres: up/up (\textit{top row}) and up/down (\textit{bottom row}) configurations. In both cases, $r_\text{ov}^\text{c} = 0.87$. 
The up/up case generally features two peaks, for observers within $45^\circ$ above and below the equatorial plane.
The up/down case, as a result of
its 4 caps, has a strong intensity observable from most directions, and features sub-peak structure, including four of almost equal strength at a viewing angle of $75^\circ$.
The magnetic field configurations for these two cases are shown in figures~\ref{fig:double_dipole} and~\ref{fig:Lines_2dip_uu}.
}
\end{figure}

We apply the SL model to predict the high-energy emission
from these two double dipole FF fields in the same way
as we did above for a  single NS.
Figure~\ref{fig:atlas_double_dipole} shows the resulting sky maps and light curves.

In the up/up case (upper panel of figure~\ref{fig:atlas_double_dipole}), sky map stagnation
occurs around two clearly distinguishable phases of stellar rotation.
As a result, light curves at every observation latitude sampled here
display double peaks of emission, which contrasts with the generally featureless
sky map and light curves of the single, aligned dipole (see panel (b) of
figure~\ref{fig:atlas_offset_aligned} and note that the $z$-offset does
not change the light curves).
It is worth stressing that double peaks are apparent for observing angles within $45^\circ$ above and below the equatorial plane.

In the up/down configuration (lower panel of figure~\ref{fig:atlas_double_dipole}), 
we again have apparent $\pi$-periodicity in $\Phi$.
In this case, however, the sky map structure is richer,
consisting of a combination of two components.
The bright region extending over a large range of latitudes
corresponds to the LOFLs extending in the direction of the angular momentum
axis of the enclosing surface. 
The other component is more concentrated near the equator.
Such a rich sky map structure translates
into light curves with as many as four peaks,
apparent with almost equal strength at $\xi_\text{obs} = 75^\circ$. 
Interestingly, the visibility in this case clearly spans a larger
range of observation angles in contrast to the up/up scenario.

Our results for the double dipole magnetospheres show that multipolar fields can be inferred from the observation of four peaks of comparable strength in gamma-ray light curves. However, multipolar fields can also produce only two main peaks, depending on the observation angle. Thus, one cannot necessarily rule out significant multipolar structure even for pulsars producing standard double-peaked light curves. Sub-peak structure appears generic in the up/down case, but even a single dipole can produce sub-peak structure (for a prominent example, see the $y$-offset inclined dipole in figure~\ref{fig:atlas_offset_inclined-y}).

%%%%%%%%%%%%%%%%%%%%%%%%
\subsection{Finding caps in binary systems}
\label{sec:Currents}
%%%%%%%%%%%%%%%%%%%%%%%%

Binary systems are not immediately amenable to the SL model, as the structure of the CS is more complex than in the single NS scenario. The shape of the CS depends on the motion of both compact objects of the binary.
For instance, the orbital motion of a magnetized NS in a binary at orbital frequency $\Omega$ (even if non-spinning) induces a CS roughly at a distance $\approx c/\Omega$ away. But additional structure in the CS can also form at smaller radii, including off the orbital plane due to the orbiting  companion~\cite{palenzuela2013electromagnetic,ponce2014interaction,Orbiting,most2020electromagnetic}.
In the particular case of BHNS binaries, this complex CS has an additional component emanating from the region near the BH 
(see, e.g., refs.~\cite{Palenzuela:2010nf,Neilsen_2011,East:2021spd,Carrasco:2021jja}).

The presence of a complex CS structure and its associated effect on the topology of the magnetic field lines makes the numerical process of finding open field lines significantly more difficult than in the single NS case. It is particularly challenging that the ``light cylinder'' notion is less precise in the binary case. As described in section~\ref{sec:emission_model}, the LC for a single NS serves to differentiate open and closed field lines.  
However, for pulsars, emission from the polar caps is also associated with intense current flow~\cite{Bai_2010}. For the binary case, we therefore seek an ``effective LC'' LC$_{\text{eff}}$ such that the associated polar cap covers some of this strong-current region. As a second guiding criterion, we also require that the polar cap does not depend sensitively on broad variations of the LC$_{\text{eff}}$. Since the effective LC is an imprecise notion, whether it is reasonable is clearly an empirical question, which we evaluate below.

\begin{figure}[!ht]
\begin{center}
\includegraphics[width=4.95cm, height= 4.95cm, frame]{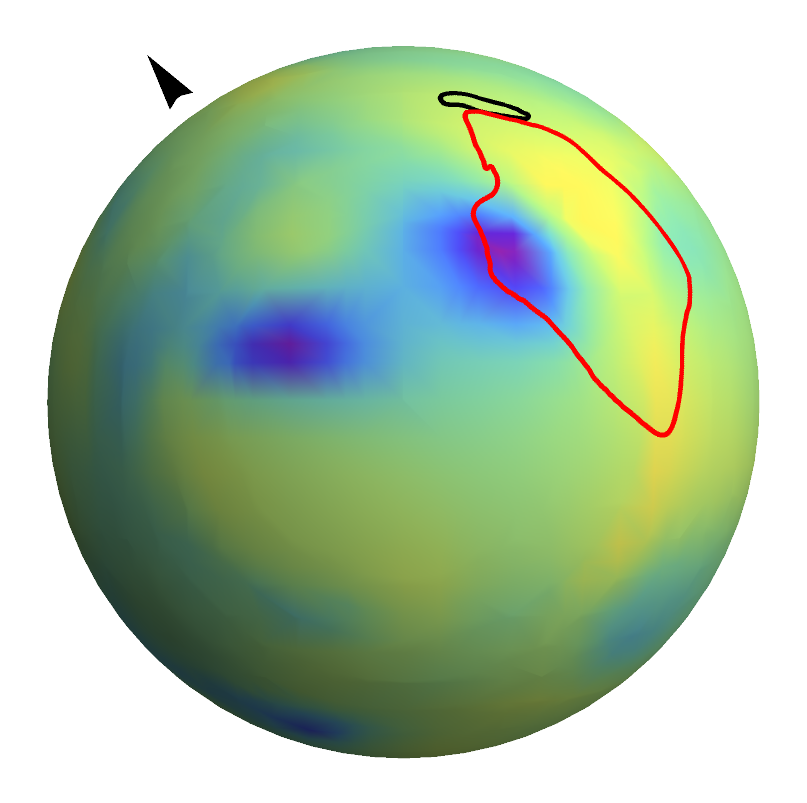}
\hspace{0.cm}
\includegraphics[width=4.95cm, height= 4.95cm, frame]{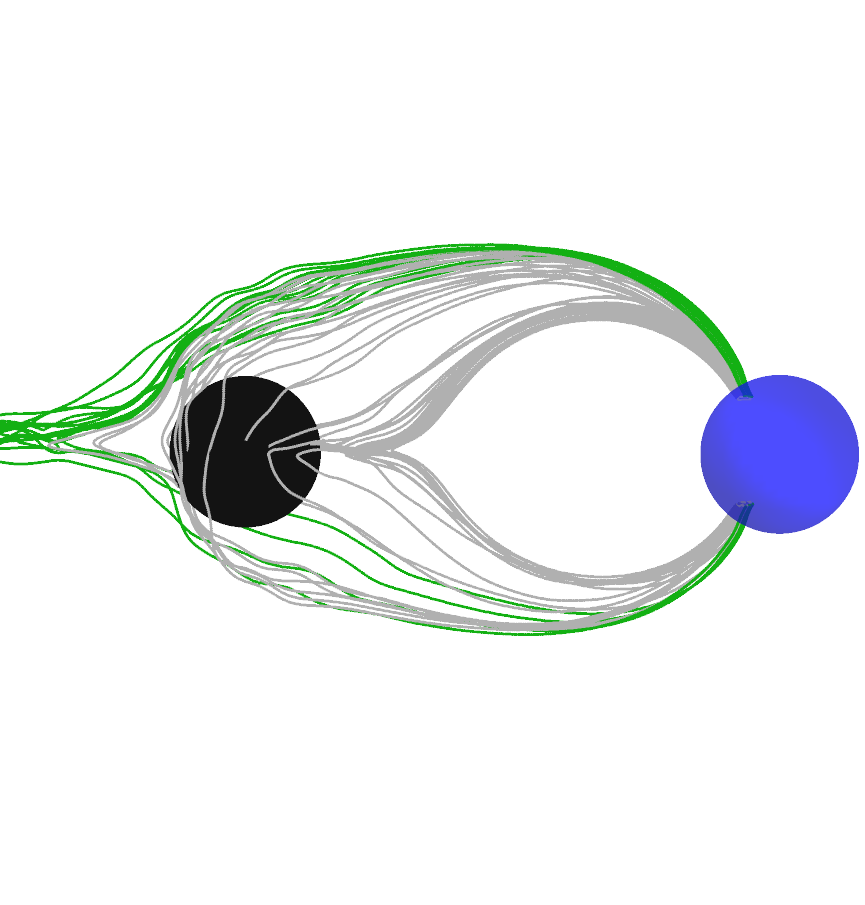}
\hspace{0.cm}
\includegraphics[width=4.95cm, height= 4.95cm, frame]{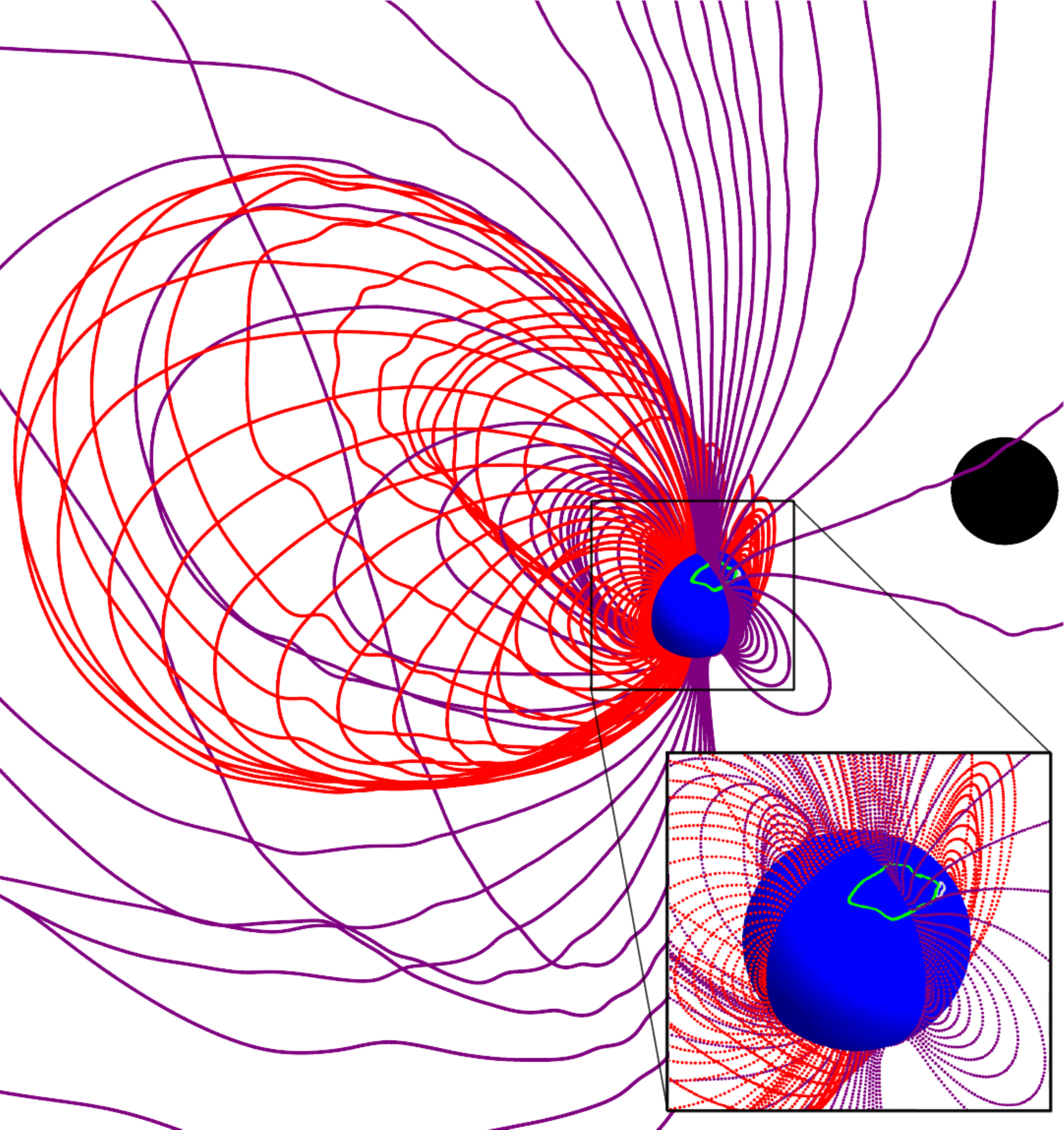}
\end{center}
\caption{\label{fig:current_BHNS} \textit{Left panel:} Northern view of the NS in a BHNS binary system at roughly $1.5$ orbits before merger. The black arrow head points in the direction of the BH component of the binary. The surface colormap is chosen to allow ease of visual comparison with past work, and signifies $\lambda$ as defined in eq.~\eqref{eq:lambda}. In-going currents (blue) and out-going currents (yellow) on the stellar surface are shown.
The southern view of $\lambda$ is approximately anti-symmetric about reflection across the equatorial plane. This means that zones of ingoing current on the northern hemisphere correspond to zones of outgoing current on the southern hemisphere (and vice-versa).
The red line represents the polar cap boundary obtained from a choice of effective LC radius $\rho_{\rm LC, eff}$ (see section~\ref{sec:Currents}) such that the cap is not sensitive to significant variations in $\rho_{\rm LC, eff}$ and the polar cap encloses strong current zones.
The black line represents the boundary of the ``BH cap'' (see section~\ref{sec:BHNS}).
The lack of orderliness in the cap, as well as the existence of more than one region of strong current flow, reflects the complex magnetic field topology and resulting CS of the binary. The intensity maps for this binary are described in section~\ref{sec:BHNS}. \textit{Center panel:} Field lines from the BH cap (gray) and from the segment of the polar cap nearest to the BH cap (green). The green field lines are LOFLs, and graze the BH before aligning with a possible equatorial current sheet. \textit{Right panel:} The magnetic field structure in the nearby region around the star. Red lines emanate from a great circle oriented orthogonal to the NS velocity. Purple lines emanate from a great circle oriented orthogonal to the BHNS separation. Field lines are generally observed to be dragged behind the NS.
}
\end{figure}

In a FF magnetosphere corotating with a NS, and ignoring curvature effects,\footnote{See Appendix A of~\cite{Gralla:2017nbw} for the definition of the field-aligned current in curved space. The primary modification is a redshift factor.} the electric current per
magnetic flux, $\lambda$, satisfies~\cite{Gruzinov:2005xp,Bai_2010} 
\begin{equation}
\lambda {\bf B} = \nabla \times \left[ {\bf B} + {\bf v} \times \left( {\bf v} \times {\bf B} \right) \right], \label{eq:lambda}
\end{equation}
where ${\bf v} = {\bf \Omega} \times {\bf r}/c$ is the corotation velocity normalized by $c$. When applying eq.~\eqref{eq:lambda} to a BHNS binary, we use a Newtonian estimate of the Keplerian orbital angular frequency ${\bf \Omega} = \Omega_{\rm Kepler}\, \hat{\bf z}$ (we also tried decreasing the frequency by more than $50\%$, and our results were insensitive). The position vector ${\bf r}$ is treated in a Newtonian sense based on the coordinates, with the origin at the center-of-mass of the binary.
Positive (negative) $\lambda$ corresponds to current flow that is parallel (anti-parallel) to the local magnetic field. We compute $\lambda$ on spheres surrounding isolated pulsars, where we know the polar cap in advance, and we reproduce results from ref.~\cite{Bai_2010} (not shown).
For the BHNS binary case, we approximate $\lambda$ on the NS surface $\approx 1.5$ orbits before merger by using eq.~\eqref{eq:lambda} on a spherical surface with slightly larger radius than the NS, and centered on the NS. The result is displayed with a colormap in the left panel of figure~\ref{fig:current_BHNS}, where blue (yellow) represents the most intense ingoing (outgoing) current zones. The orbital angular frequency of the binary at this instant is approximately $\Omega_{\rm Kepler} \simeq 1740$ rad/s. 
The map of $\lambda$ in the left panel of figure~\ref{fig:current_BHNS},
although anti-symmetric with respect to reflections about the equatorial plane, contains regions of strong current flow. Despite the initialization of the magnetic field as an aligned dipole, one of the strong current regions is an elongated shape (yellow), which is reminiscent of single pulsars with intermediate dipole inclinations reported in past work (see figure 4 from ref.~\cite{Bai_2010}, inclinations $30^\circ$ \& $60^\circ$).\footnote{Significant current regions were also reported outside the polar cap in the $30^\circ$ inclination case in figure 4 from ref.~\cite{Bai_2010}, just as we observe in the left panel of figure~\ref{fig:current_BHNS}.}

Nonetheless, empirically we find a reasonable LC$_{\rm eff}$ that produces a cap covering two regions of strong ingoing and outgoing current. This LC has a radius $\rho_{ \rm LC, eff}$ equal to roughly 8.3 times the instantaneous orbital separation of the binary $R_{\rm bin}$. This empirical polar cap is displayed as a red solid line in the left panel of figure~\ref{fig:current_BHNS} and as a green solid line in the right panel. We find broad insensitivity of the cap to choices of effective LC radii in the range $\rho_{ \rm LC, eff}\approx (4-14)\times R_{\rm bin}$. For reference, the naive choice of light cylinder radius based on a Newtonian estimate of the orbital angular frequency would be $c/\Omega_{\rm Kepler} \approx 2.4 \times R_{\rm bin}$.
Given this polar cap, a pragmatic application of the SL model for high-energy EM radiation then follows just as in the single pulsar scenario.

%%%%%%%%%%%%%%%%%%%%%%%%%%%%%%%%%%%%%%%%%%%%%%%
%%%%%%%%%%%%%%%%%%%%%%%%%%%%%%%%%%%%%%%%%%%%%%%
\subsection{Black hole--Neutron star binary}
\label{sec:BHNS}
%%%%%%%%%%%%%%%%%%%%%%%%%%%%%%%%%%%%%%%%%%%%%%%
%%%%%%%%%%%%%%%%%%%%%%%%%%%%%%%%%%%%%%%%%%%%%%%

Now we turn our attention to the case of a BHNS binary. In such a system, the CS structure is more complex than even a BNS system, which makes a straightforward application
of the SL model difficult. In addition to the usual open and closed field lines, the subset of lines entering the BH have an ambiguous characterization, since any current which flows into the BH will neither return to the star nor leave the system. It has also been observed in simulations~\cite{East:2021spd,Carrasco:2021jja} that the BH induces reconnections in its vicinity, which
give rise to new, possibly disconnected contributions to the CS in the BH's ``wake.'' Such phenomena are likely to cause additional high-energy 
emission (see e.g.~\cite{Crinquand:2020reu}).
A method which attempts to determine the CS based solely on tracing field lines from large distances will likely fail to capture this additional emission region. Furthermore, even if the CS lies in the equatorial plane, the orbital motion of a magnetized NS itself causes the CS to exhibit a spiral-shaped inner edge. This spiral contrasts with the single pulsar case, where the inner edge of the CS\footnote{That is, where the CS transitions to thicker current layers, see e.g.~\cite{Bai_2010}.} is circular. It is also known that plasmoids in the BH's vicinity can produce further contributions that
must be accounted for~\cite{most2020electromagnetic,East:2021spd}.
We defer the development of strategies to account for these various missing emissions to future work. Here we obtain sky maps
associated only with the large-scale magnetic field behavior, as captured by the LOFLs and magnetic polar caps.

We study the specific case of an irrotational BHNS binary ---meaning that the binary constituents lack any
initial intrinsic spin--- with approximately 5:1 mass ratio ($M_\text{BH}= 7 M_{\odot}, M_\text{NS} = 1.33 M_{\odot}$). 
The star's initial magnetic field is a centered dipole aligned
with the orbital angular momentum. This system has been simulated in ref.~\cite{East:2021spd} with full
general relativity and resistive magneto-hydrodynamics for roughly $2.5$ orbits through merger.
Here, we analyze a snapshot from this simulation roughly $1.5$ orbits before merger, at which time transients from the initialization of the magnetic field have already propagated away from the system.
We employ the SL model for gamma-ray emission with LOFLs and magnetic polar caps determined by
the method described in section~\ref{sec:Currents}---see also figure~\ref{fig:current_BHNS}. As a sanity check for our polar caps,
we compute the induced current per magnetic flux $\lambda$, as defined in eq.~\eqref{eq:lambda}, on a sphere slightly larger than the NS which completely encloses the star. Computing $\lambda$ on a sphere that is slightly larger than the star is a practice followed in ref.~\cite{Bai_2010}.\footnote{Such a choice of extraction surface ensures that $\lambda$ is not computed inside dense material where eq.~\eqref{eq:lambda} does not hold.} We see in figure~\ref{fig:current_BHNS} that the caps enclose significant ``hot spots'' of positive and negative current. However, the same figure also indicates non-negligible hot spots outside of the polar cap. Although such additional hot spots have been observed with single pulsars in previous work,\footnote{See figure 4, 30 degree inclination, in ref.~\cite{Bai_2010}.} we cannot rule out that the extra contributions are due to the rapid orbital motion of the NS, a recent reconnection event in the vicinity of the binary, or the BH and its effect on nearby field lines. We visualized field lines emanating from the most prominent hot spot residing outside the polar cap in figure~\ref{fig:current_BHNS}, and found them to be unremarkable and unambiguously closed. Also note that $\lambda$ was computed using the estimated Keplerian orbital angular frequency $\Omega_{\rm Kepler} \approx 1740$ rad/s, whereas the frequency implied by our choice of effective LC is about 60\% smaller. However, we checked that our results for $\lambda$ are not sensitive to significant changes in the frequency used in eq.~\eqref{eq:lambda}.

In the left panel of figure~\ref{fig:current_BHNS}, we display the northern hemisphere of the star. The colormap represents $\lambda$ on a linear scale, and the southern hemisphere appears anti-symmetric (i.e.~with $\lambda \rightarrow -\lambda$). The BH is situated in the direction indicated by the arrowhead. The magnetic polar cap is displayed as a red line in the left panel, and as a green line in the right panel. We also draw what we call the ``BH cap'' as a black line in the left panel (gray line the right panel). The BH cap encloses those field lines which enter the BH. The BH cap is remarkably close to the polar cap, nearly sharing a segment of its boundary. Field lines which emanate from points that are squished between the BH cap and polar cap are the only closed field lines which pass over the BH. By ``passing over'' the BH (or ``under'' for the southern hemisphere), we mean in the sense of reaching the far side of the BH (leftmost region in the middle panel of figure~\ref{fig:current_BHNS}). This indicates that not many field lines pass over the BH and return to the star; most field lines that pass over the BH are open. This field line behavior is illustrated in the center panel of figure~\ref{fig:current_BHNS}, where field lines have been integrated from the BH cap (gray), as well as from the segment of the magnetic polar cap which nearly borders the BH cap (green). 

The proximity of the polar and BH caps can be understood in terms of the location of CS structures. We observe in the middle panel of figure~\ref{fig:current_BHNS} that field lines near the BH on the equatorial plane tend towards a parallel orientation with that plane. This is an indication that a CS is present (or forming) on the equatorial plane near the BH, which thereby supports a discontinuity in the component of the magnetic field that is parallel with the equatorial plane. CS structures induced by the BH in its immediate vicinity have been reported in past results~\cite{East:2021spd,Carrasco:2021jja}. It is also worth noting that, near merger, the length scale of the orbital LC ($c/\Omega_{\rm Kepler}$) is only larger than the orbital separation by a factor of order $\mathcal{O}(1)$. This means that the CS induced by the orbital motion of the magnetized NS is expected to be relatively close to the BH, especially given the spiral-shaped inner edge of the CS. The system near the BH is therefore quite crowded with CS structures, thereby preventing most field lines which pass over the BH from returning to the star. Thus, one can generically expect that the BH cap and the magnetic polar cap sit very close to each other for BHNS binaries near merger, as displayed in the left panel of figure~\ref{fig:current_BHNS}.

The field line structure in the vicinity of the star is displayed in the right panel of figure~\ref{fig:current_BHNS}. We refer to regions left, right, in front, and behind the star with respect to an observer standing vertically on the star in the right panel of figure~\ref{fig:current_BHNS} and facing parallel with the star's velocity. Two sets of field lines are shown: a red set, which is integrated from a great circle on the star whose plane is orthogonal to the star's velocity, and a purple set, which is integrated from a great circle that is coplanar with both the star's velocity and the orbital angular momentum. If the star were stationary and isolated, the red field lines would not preferentially occupy the regions in front or behind the star. Instead, in the binary system the red field lines preferentially occupy a region behind the star. This indicates a relative deficit of field lines on the left and right sides of the star.\footnote{By ``deficit'' we mean that, if one maps a volume to the surface of the star via field lines, the surface area of the image of that map is smaller than it would otherwise be in the absence of the orbital motion. This is also the sense in which we talk about the ``density'' of field lines.} The purple field lines display an even more significant deficit of field lines in front of the star, with a higher density of lines behind the star. Both sets of field lines indicate that the density of field lines near the star is primarily affected by the translational motion of the NS, as though the star is dragging field lines along as it orbits the center of mass of the system. It stands to reason that this dragging morphology would imprint upon the sky map. Indeed, one can see in the right panel of figure~\ref{fig:current_BHNS} that the polar cap position is biased towards the front of the star, as opposed to being aligned with the vertical direction. This is in spite of the fact that the star's magnetic field is initialized as a centered dipole aligned with the orbital angular momentum. We interpret this forward shift of the polar caps as a direct result of the dragged field line morphology.

\begin{figure}[!ht]
\begin{center}
\includegraphics[width=17cm]{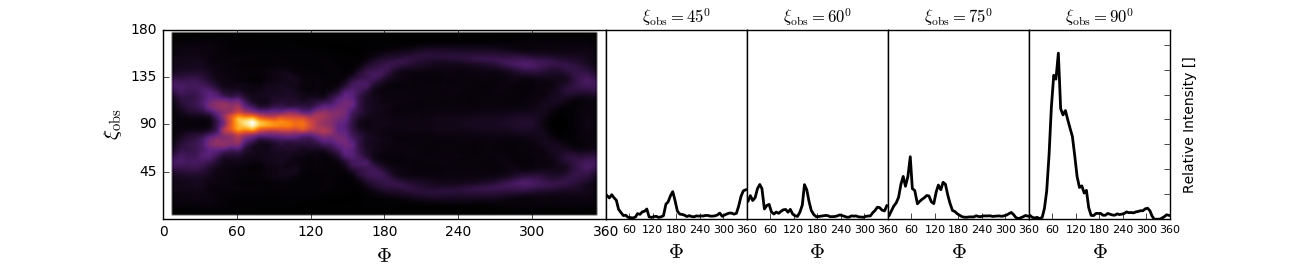}
\end{center}
\caption{\label{fig:atlas_BHNS} Sky map and light curves for the FF magnetosphere of a BHNS binary system, produced using an effective LC 8.3 times larger than the instantaneous orbital separation of the binary, and $r_\text{ov}^\text{c} = 0.87$. 
The emission zone extends up to $1.5\rho_\text{LC}$, the same as for all sky maps computed in this work.}
\end{figure}

We present the sky map for the BHNS binary in figure~\ref{fig:atlas_BHNS}, based upon the SL model applied to the magnetic polar caps. The sky map exhibits reflection symmetry about the equatorial plane, as one would expect. The emission is dominant in the equatorial plane for phases $\Phi \approx 60^{\circ}$-$120^{\circ}$, and appears to deflect to nearly vertical directions $\xi_{\rm obs} \approx 150^\circ$ at a phase $\Phi\approx 200^\circ$. Consequently, observers whose viewing angles are neither equatorial nor vertical will generally see 2 peaks (see the right panel of figure~\ref{fig:atlas_BHNS} for $\xi_{\rm obs}\in \lbrace 45^\circ, 60^\circ, 75^\circ \rbrace$). Face-on observers ($\xi_{\rm obs}\in \lbrace 0^\circ, 180^\circ \rbrace$) receive little emission, and equatorial observers primarily see one broad peak (see the right panel of figure~\ref{fig:atlas_BHNS} for $\xi_{\rm obs} = 90^\circ$). 

\begin{figure}[!ht]
\begin{center}
\includegraphics[width=17cm]{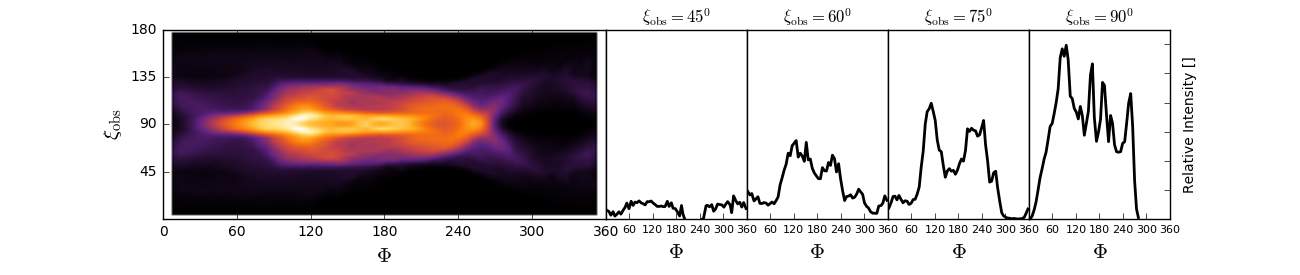}
\end{center}
\caption{\label{fig:sky_offset_x9} Sky map and light curves for an $x$-offset dipole with $d/R = 0.9$. In section~\ref{sec:BHNS} we compare this sky map with that from the BHNS binary shown in figure~\ref{fig:atlas_BHNS}. The central open volume coordinate is $r_\text{ov}^\text{c} = 0.87$.}
\end{figure}

On the basis of the enclosing surface approximation, one might expect the BHNS sky map to resemble that of the aligned, offset dipole, or the ``orbiting NS'' we considered previously (see figure~\ref{fig:atlas_oNS}), since the magnetic field is sourced by a single object localized at a position offset from the system's center of mass. However, in those cases we do not observe significant deflection of emission from equatorial to nearly vertical directions, which leaves a dark equatorial region on the sky map in figure~\ref{fig:atlas_BHNS} (see the range $\Phi\approx 180^\circ$-$\, 360^\circ$). In an effort to obtain qualitative features of the BHNS sky map in the enclosing surface approximation, we analyze an aligned, offset dipole with a more extreme offset than in figure~\ref{fig:atlas_offset_aligned} ($d/R=0.9$ rather than $d/R=0.6$). With the larger offset, a dark equatorial region emerges, reminiscent of the BHNS sky map---see the left panel of figure~\ref{fig:sky_offset_x9}, between $\Phi\approx 280^\circ$-$\, 360^\circ$. Notably, the light curves for $\xi_{\rm obs}=60^\circ$ and $\xi_{\rm obs}=75^\circ$ also exhibit a similar shape (the ``cat ears'' shape, consisting of two peaks separated by moderate amplitude emission). The intense curvature near the BH horizon, together with the irrotational character and rapid translational motion in the BHNS binary, arguably transforms the emission pattern in significant ways, possibly accounting for the differences with the $d/R=0.9$ offset aligned dipole.  

\begin{figure}[!ht]
\begin{center}
\includegraphics[width=7.5cm, frame]{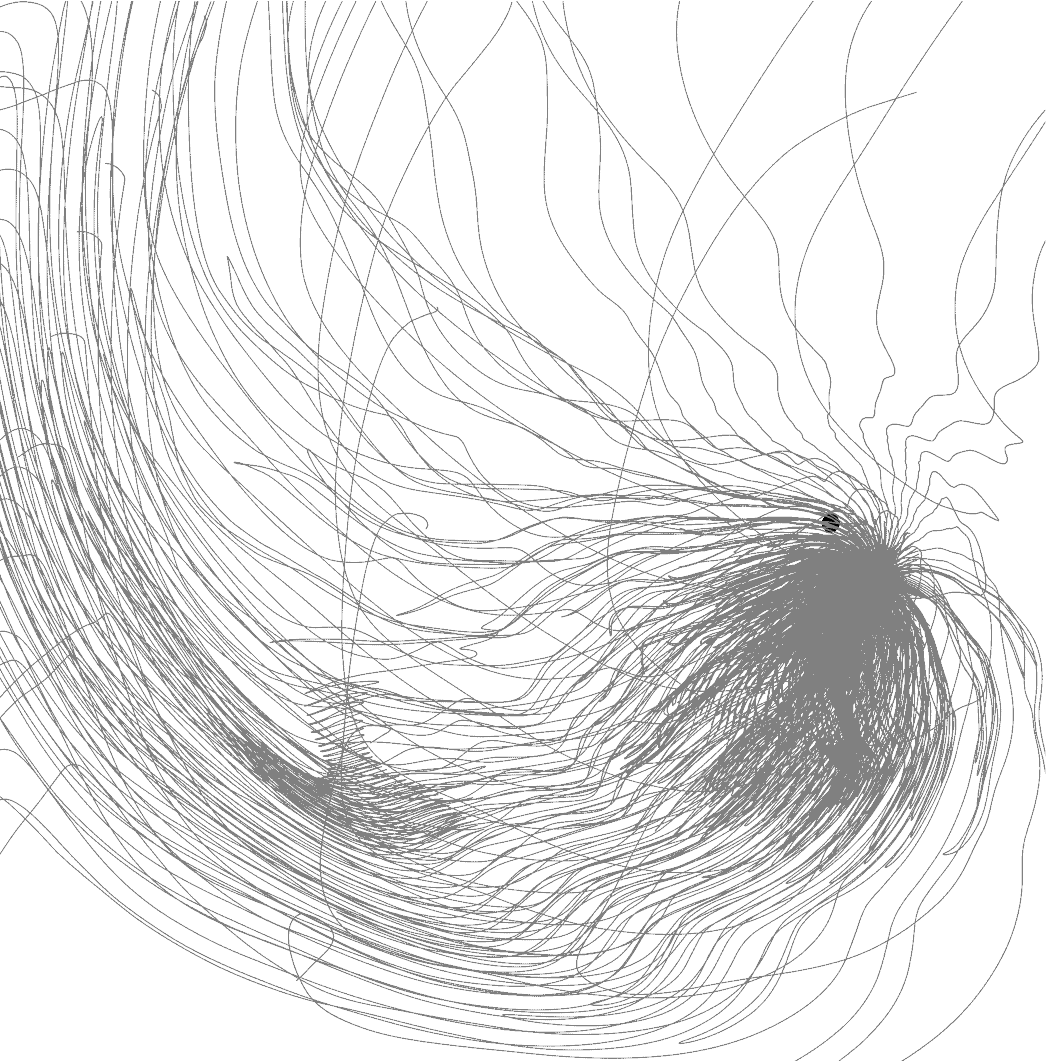}
\end{center}
\caption{\label{fig:BHNS_lines_TopView} Equatorial plane of the BHNS binary system described in section~\ref{sec:BHNS}, as it orbits counter-clockwise. The BH is visible, whereas the NS is hidden behind gray magnetic field lines, which display a winding morphology. These lines come from seeds on the star surface placed at 90 (45) uniformly distributed values of the azimuthal (polar) coordinate.}
\end{figure}

A natural question is which value of $\Phi$ corresponds to a ``BH eclipse'' in figure~\ref{fig:atlas_BHNS}, i.e.~when the BH is geometrically behind the star in a coordinate sense. The BH eclipse corresponds to roughly $\Phi = 308^\circ$, which is a somewhat unremarkable phase of the sky map. The NS eclipse is $180^\circ$ away from that, at $\Phi=128^\circ$. However, it is important to note that the field lines exhibit a winding morphology, as though they are being dragged by the NS motion. The winding field line morphology is displayed in figure~\ref{fig:BHNS_lines_TopView}, where the binary is shown in its orbital plane as it orbits counter-clockwise, the BH is visible, and the star is hidden behind gray field lines. The field lines therefore appear to encode the recent history of the NS motion, and thus the sky map will lag in phase $\Phi$ with respect to the binary. However, the geometric eclipses are very indirect ways to understand the sky map because the emission regions exist along the separatrix layer, which lies away from the compact objects. A more meaningful correspondence between the sky map and the binary would involve a mapping from points on the sky map to the volumes where the photons in question had originated. Such a notion was conveyed in the single pulsar case in figures 6 \& 8 in ref.~\cite{Bai_2010}, for example. One could also imagine a volume rendering, whereby emitting volumes in the system are color-coded in correspondence with the regions of the sky map that their emitted photons contribute to. Such a visualization is beyond the scope of this work.

A caveat for the BHNS light curves is that the orbital parameters are changing on a time scale of an orbit. Thus, the sky map and light curves in figure~\ref{fig:atlas_BHNS} that we derived from a snapshot in time do not actually represent what any observer would see over time. In other words, the light curves are presented as functions of viewing phase $\Phi$, not as functions of time. In order to construct the light curves in time, one would need to piece together several snapshots over an orbital period or more. Furthermore, to the extent that the field line structure is not stationary, the application of eq.~\eqref{eq:lambda} when defining $\lambda$ in the left panel of figure~\ref{fig:current_BHNS} will be inaccurate (in addition to inaccuracies stemming from the application of eq.~\eqref{eq:lambda} in a curved spacetime).

It is difficult to assess the expected absolute strength of gamma-ray emission from the BHNS binary. A reasonable order-of-magnitude upper bound can be obtained by considering the different components that would contribute to the overall budget of energy output. A spinning NS can yield $L_s \sim c^{-3} B^2 R^6 \Omega_s^4$, and the orbital motion of the binary can yield $L_b \sim c^{-3} B^2 R^6 \Omega_b^6 d^2$, where $B$ is the surface magnetic field strength, $R$ is the stellar radius, $\Omega_s$ is the spin frequency of the star, $\Omega_b$ is the orbital frequency of the binary, and $d$ is the orbital separation.
The spin contribution is expected to be small compared to the orbital component in binary mergers, since EM spin-down would have significantly reduced $\Omega_s$ in comparison with $\Omega_b$. And the binary contribution can be quite significant, especially
at late stages of the inspiral. Assuming
a surface magnetic field strength of $10^{11-12}$ G, luminosities of the order of $10^{40} \mbox{ erg s}{}^{-1}$ can be achieved at merger based on unipolar induction 
estimates~\cite{Hansen:2000am,mcwilliams2011electromagnetic}.\footnote{Importantly, full GR-RMHD simulations reveal that a larger peak value is achieved; a sustained strong emission is possible in the last few orbits of the binary~\cite{East:2021spd,Carrasco:2021jja}, and an enhanced final output occurs as coalescence takes place~\cite{East:2021spd}.}  
Under optimistic assumptions about the efficiency of gamma-ray production, detection out to $\approx 100$ kpc with the Fermi Gamma Ray Space Telescope would be possible 
(see e.g.~ref.~\cite{10.1093/ptep/ptaa126}).
Most detected gamma-ray pulsars are galactic (e.g.~\cite{Saz_Parkinson_2010,doi:10.1146/annurev-astro-081913-035948}), though detection at farther distances is achieved by raising the detection significance with long time integrations~\cite{PSRGAMMA801}. It is therefore pertinent to note that the significance of an otherwise tenuous detection of gamma-ray photons from a compact binary would be increased if they were coincident with the timing and directionality of a GW signal. Long time integrations should also take into account the decaying orbital period of the binary. Nevertheless, compact binary merger rates imply that gamma-ray detection from such systems would be improbable.

%

%%%%%%%%%%%%%%%%%%%%%%%%%%%%%%%%%%%%%%%%%%%%%%
%%%%%%%%%%%%%%%%%%%%%%%%%%%%%%%%%%%%%%%%%%%%%%
\section{Discussion}
\label{sec:Discussion}
%%%%%%%%%%%%%%%%%%%%%%%%%%%%%%%%%%%%%%%%%%%%%%
%%%%%%%%%%%%%%%%%%%%%%%%%%%%%%%%%%%%%%%%%%%%%%

Understanding the high-energy EM emission from NSs, both in isolation and in binaries, has become increasingly important in the age of multimessenger astronomy. In this work, we have presented sky maps and light curves for gamma-ray emission 
from many variations of magnetic field configurations built from dipoles. We have been motivated by recent x-ray observations by NICER---which 
suggest that the magnetic field of NSs could generally be more complicated than a single, centered dipole---as well as potential EM precursor signals 
from merging BNS and BHNS systems, which are targets for gravitational wave detection by LIGO/VIRGO~\cite{TheLIGOScientific:2017qsa, Abbott_2021}.  To this end, we have extended Bai \& Spitkovsky's  Separatrix 
Layer model of gamma-ray emission \cite{Bai_2010} to investigate these scenarios.

We have performed FF simulations of NS magnetospheres, starting from the simplest scenario of an isolated pulsar and then gradually incorporating relevant features that might serve as guidance---or even approximations---to the more complex magnetospheres of compact binary systems. By applying the SL model to these numerical solutions, we are able to predict at least a subset of the gamma-ray emission.
We have also discussed the required modifications of the algorithm presented in ref.~\cite{Bai_2010} when applying the SL method to non-vacuum binaries. In particular, the complexity of the current sheets (e.g.~refs.~\cite{palenzuela2013electromagnetic,Orbiting, most2020electromagnetic,East:2021spd,Carrasco:2021jja}) prevents a direct application of the original method, and leads to uncertainties in identifying the emission regions. In the context of binaries, we argue that the SL model therefore accounts for the subset of the emission associated with the large-scale CS, which is typically induced by the orbital motion of magnetized stars.
In this work, we have taken a first step towards addressing these issues and presented the first sky maps of gamma-ray emissions from a BHNS binary close to merger.

We have used a standard isolated pulsar magnetosphere, with a centered magnetic dipole and with a few inclination angles, to validate our approach by comparing with previous results (in particular, those of ref.~\cite{Bai_2010}).
These same sky maps and light curves serve as a reference for comparing with the more complicated configurations studied in this work.

We introduced a twist in the magnetosphere, following a simple model motivated by crustal deformations in magnetars (e.g., ref.~\cite{parfrey2013}). Our results  show that stable configurations (for under-critical twists) are indeed possible within the closed-zone of the rotating NS, and that such current-carrying bundles would not significantly affect the gamma-ray sky maps as long as they do not intersect the separatrix layer. Similar confined magnetospheric currents have been recognized, for instance, along the flux-tube that connects the BH with the NS in BHNS binaries.

We then considered the effects of dipoles not concentric with the star. Such configurations have several astrophysical motivations, and indeed off-setting the dipole from the center significantly changes the predicted sky map. In particular, offsets within the equatorial plane break the symmetries of the sky map quite differently from a vertical offset, and thus, give hope that such configurations can be differentiated with observations as they would typically produce two peaks with different amplitudes in their light curves.

We have introduced a new technique, which we call the {\it enclosing surface approximation}, in which a fictitious perfectly conducting sphere encloses a BNS (two dipoles) or a BHNS (single dipole) system.
This approximation seeks to model the large-scale magnetic field far from the binary without having to model the extreme dynamics of the binary itself. Our initial results suggest the promise of such a model that we have tested by direct comparison with a single orbiting NS setting from ref.~\cite{Orbiting}.
We applied the approximation to double dipole settings, both equally displaced and with the same strength, for both aligned (up/up) and anti-aligned (up/down) configurations. We found that the up/up case generically leads to double-peaked emission, spanning an observation range of roughly $45^\circ$ about the equator. Whereas the up/down case showed richer structure both in the magnetic field topology and on its sky map, which involves a much broader range of observation angles and produces light curves with as many as four emission peaks. These results---and similar ones, exploring different parameters---will provide valuable information to understand high-energy emissions from BNS systems.

Finally, we applied the SL model to a snapshot of the numerical evolution of a BHNS system in full general relativity. This application required 
careful modifications to the method
for finding open field lines, made more involved by the complicated CS produced by the dynamics of the binary. The resulting sky map showed a broad intensity peak within roughly $30^\circ$ around the orbital plane. Thus, an equatorial observer will detect a single broad peak within one orbital period of the binary. For more generic viewing angles, observers will see double-peaked emission, although much weaker signals than those received by near equatorial observers. On the other hand, very little emission is expected towards the polar regions. 

We can speculate about the pre-merger gamma-ray emission from the BNS merger GW170817. If the BNS was similarly magnetized as the BHNS system we studied here (i.e.~one star magnetized with an aligned dipole), then its sky map might also have had a deficit of polar emission relative to equatorial directions. If so, then the pre-merger gamma-ray signal would have been faint  (relative to edge-on views) since GW170817 is estimated to have been viewed nearly face-on (at most $\xi_{\rm obs}\approx 30^\circ$, see e.g.~\cite{mandel2018orbit}).

Based on recent estimates~\cite{Carrasco:2021jja}, even in the most favorable BHNS scenarios with high BH spins ($\gtrsim 0.9$), for which $L_{\rm EM} \approx 10^{42-46} \, [B_p/ 10^{12}{\rm G}]^2  \, \mbox{ erg s}{}^{-1}$,  the detection prospects for EM signals in the high-energy band by current facilities (such as the Swift Alert Telescope or the Fermi Gamma-ray burst monitor) are not very optimistic for extra-galactic sources at $\gtrsim 100$\,Mpc. On the other hand, for sources in our galaxy and nearby, the features identified here can be used to help characterize the magnetic field structure of gamma ray pulsars. 

A number of caveats accompany this work. First among these is general uncertainty in identifying the proper emission region for each system. One standard approach for a single star is the determination of a polar cap of open field lines, but the magnetic field configuration for double dipoles and binaries can be very complicated, making a determination of open regions much more difficult. Secondly, we have not included gravitational effects on the photon trajectories, which can introduce modulations
in the BHNS system case, especially near merger. Finally, we have studied just a few configurations, but the parameter space of magnetic field configurations is vast and largely unconstrained by both observation and theory. In this regard, the enclosing surface approximation could be a valuable tool to systematically explore the parameter space and provide the main qualitative properties of the resulting high-energy emissions. Future work will aim to address some of these open issues.

%%%%%%%%%%%%%%%%%%%%%%%%%%%
%                 ACKNOWLEDGMENTS
%%%%%%%%%%%%%%%%%%%%%%%%%%%

\acknowledgments
We are happy to thank Xue-Ning Bai, Cole Miller, Anatoly Spitkovsky, Carlos Palenzuela, and Huan Yang for interesting discussions.
This work was supported by the NSF under grants PHY-1827573, PHY-1912769, and PHY-2011383 (SLL), CONACyT grants ``Ciencia de Frontera" 140630 and 376127 (NO),
as well as NSERC through a Discovery grant and CIFAR (LL),
and by the UNAM-PAPIIT grant IA100721 (NO).
This research was supported in part by Perimeter Institute for Theoretical Physics. Research at Perimeter Institute is supported by the Government of 
Canada through the Department of Innovation, Science and Economic Development Canada and by the Province of Ontario through the Ministry of Research, Innovation and Science.
Computations were performed with XSEDE computational resources, on the Yamazaki cluster from the Max Planck Institute for Gravitational Physics, Potsdam, and on the Niagara supercomputer at the SciNet HPC Consortium. 
SciNet is funded by: the Canada Foundation for Innovation; the Government of Ontario; Ontario Research Fund - Research Excellence; and the University of Toronto.

%%%%%%%%%%%%%%%%%%%%%%%%%%%%%%%%%%%%%%%%%%%%%
\bibliographystyle{JHEP}
\bibliography{skymaps}
%%%%%%%%%%%%%%%%%%%%%%%%%%%%%%%%%%%%%%%%%%%%%

\end{document}